\documentclass[useAMS,usenatbib]{mn2e}

\def\Zsol{\hbox{Z$_{\odot}$}}
\def\Msol{\hbox{M$_{\odot}$}}
\def\Lsol{\hbox{L$_{\odot}$}}

\usepackage{epsfig,natbib,url,graphicx}
\usepackage{lscape,amsmath,amssymb}
\usepackage{booktabs,threeparttable}
\usepackage{color}
\usepackage{soul} 
\usepackage{fancyvrb}
\usepackage{pifont}
\newcommand{\cmark}{\ding{51}}%
\newcommand{\xmark}{\ding{55}}%

\newcommand{\hi}{H\,{\sc i}}
\newcommand{\hii}{H~{\sc ii}}
\newcommand{\hei}{He~{\sc i}}
\newcommand{\heii}{He~{\sc ii}}
\newcommand{\kms}{km\,s$^{-1}$}
\newcommand{\eld}{$N_{\rm e}$}

\newcommand{\elt}{$T_{\rm e}$}

\newcommand{\Sp}{S$^+$}
\newcommand{\Spp}{S$^{2+}$}

\newcommand{\op}{O$^+$}
\newcommand{\nep}{Ne$^+$}
\newcommand{\opp}{O$^{2+}$}

\newcommand{\np}{N$^+$}

\newcommand{\foiii}{[O~{\sc iii}]}

\newcommand{\foi}{[O~{\sc i}]}
\newcommand{\foii}{[O~{\sc ii}]}
\newcommand{\fsii}{[S~{\sc ii}]}
\newcommand{\fsiii}{[S~{\sc iii}]}

\newcommand{\fnii}{[N~{\sc ii}]}

\newcommand{\fneiii}{[Ne~{\sc iii}]}

\newcommand{\ffeii}{[Fe~{\sc ii}]}
\newcommand{\ffeiii}{[Fe~{\sc iii}]}

\newcommand{\feii}{Fe~{\sc ii}}

\newcommand{\hp}{H$^+$}
\newcommand{\hep}{He$^+$}
\newcommand{\hepp}{He$^{2+}$}
\newcommand{\ha}{H$\alpha$}
\newcommand{\hb}{H$\beta$}
\newcommand{\hg}{H$\gamma$}
\newcommand{\hd}{H$\delta$}
\newcommand{\Lya}{Ly$\alpha$}

\title[A Chemodynamical Study of Haro~11]{The Lyman break analogue Haro~11: spatially resolved chemodynamics with VLT FLAMES\thanks{Based on observations made with ESO telescopes at the Paranal Observatory under
programme 083.B-0336A.}}
\author[B. L. James et al.]{B.~L. James$^{1,2}$\thanks{E-mail: bjames@ast.cam.ac.uk},
Y.~G.~Tsamis$^{3}$\thanks{E-mail: ygtsamis@gmail.com},
J.~R.~Walsh$^{3}$,
M.~J.~Barlow$^{4}$,
M.~S.~Westmoquette$^{3}$
\\
$^{1}$Space Telescope Science Institute, Baltimore, MD 21218\\
$^{2}$Institute of Astronomy, University of Cambridge, Madingley Road, Cambridge, CB3 0HA, UK\\
$^{3}$European Southern Observatory, Karl-Schwarzschild Strasse 2, D-85748 Garching bei
M\"unchen, Germany\\
$^{4}$Department of Physics and Astronomy, University College London, Gower Street, London, WC1E 6BT
}

\begin{document}

\date{Accepted 2013 January 4.  Received in original form 2012 October 18}

\pagerange{\pageref{firstpage}--\pageref{lastpage}} \pubyear{2012}

\maketitle

\label{firstpage}

\begin{abstract}
Using VLT/FLAMES optical integral field unit observations, we present the first spatially resolved spectroscopic study of the well known blue compact galaxy Haro~11, thought to be a local analogue to high redshift Lyman Break Galaxies. Haro~11 displays complex emission line profiles, consisting of narrow (FWHM $\lesssim$ 200~\kms) and broad (FWHM $\sim$ 200--300~\kms) components. We identify three distinct emission knots kinematically connected to one another. A chemodynamical analysis is presented, revealing that spatially resolved ionic and elemental abundances do not agree with those derived from integrated spectra across the galaxy. We conclude that this is almost certainly due to the surface-brightness-weighting of electron temperature in integrated spectra, leading to higher derived abundances. We find that the eastern knot has a low gas density, but a higher temperature (by $\sim$4,000 K) and consequently an oxygen abundance $\sim$0.4 dex lower than the neighbouring regions. A region of enhanced N/O is found specifically in Knot C, confirming previous studies that found anomalously high N/O ratios in this system. Maps of the Wolf-Rayet feature at 4686\,\AA\ reveal large WR populations ($\sim$900--1500 stars) in Knots A and B. The lack of WR stars in Knot C, combined with an age of $\sim$7.4~Myr suggest that a recently completed WR phase may be responsible for the observed N/O excess. Conversely, the absence of N-enriched gas and strong WR emission in Knots A and B suggests we are observing these regions at an epoch where stellar ejecta has yet to cool and mix with the ISM. 

\end{abstract}
\VerbatimFootnotes


\section{Introduction}
Blue Compact Galaxies (BCGs), due to their high star-formation rates and low metallicities, are thought to be the nearest analogues to young, starbursting galaxies at high redshift.  However, unlike their distant counterparts, the proximity of BCGs provides us with an opportunity to study galaxy evolutionary processes in great detail.  

\begin{figure*}
\begin{center}
\includegraphics[scale=.5]{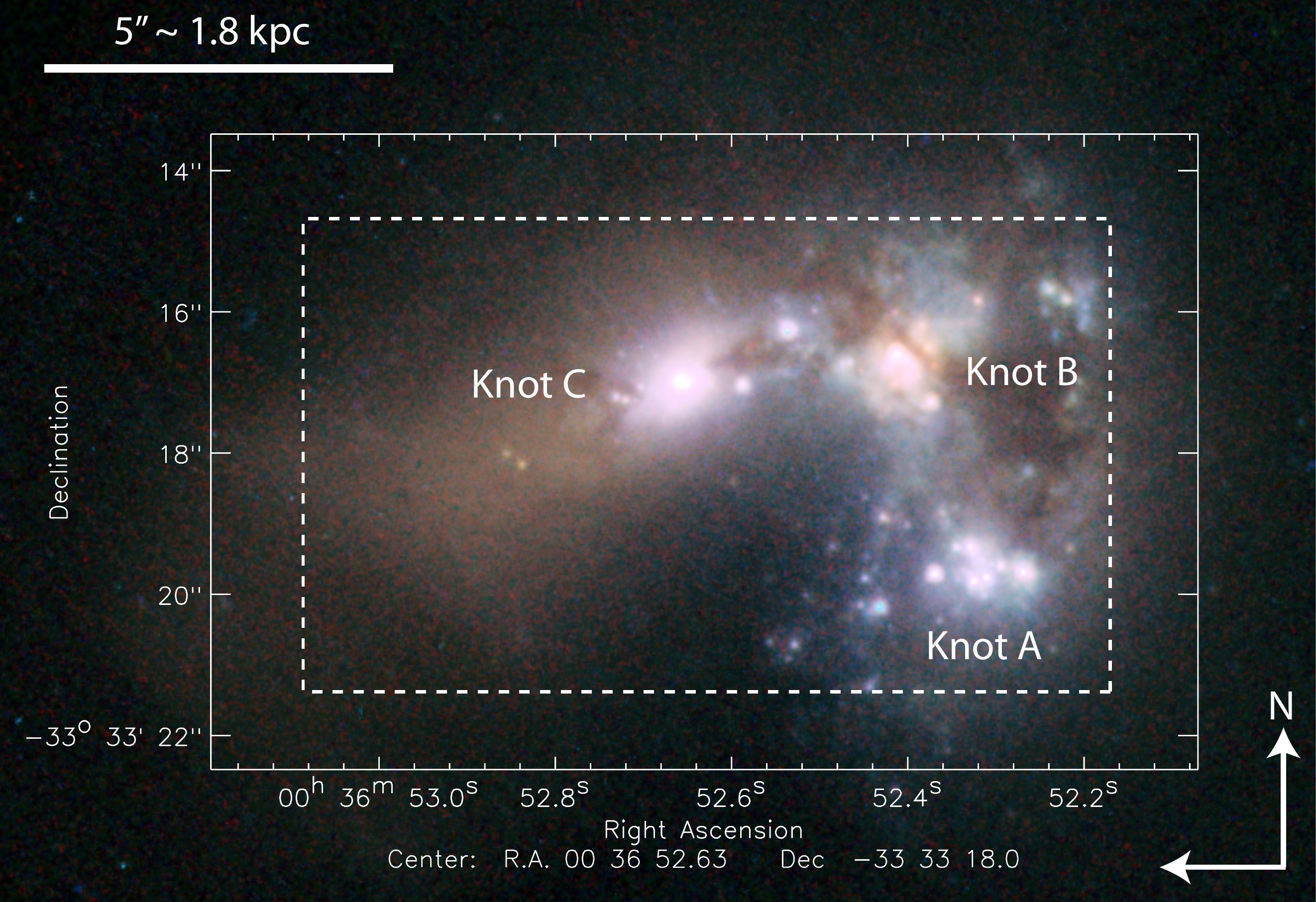}
\caption{\textit{HST} colour composite image of Haro~11 (consisting of ACS-HRC F220W, F330W and F814W along with ACS-WFC F435W and F550W images) overlaid with the 11\farcs5 $\times$ 7\farcs3 FLAMES IFU aperture (dashed line).  Knots are labeled following the nomenclature of \citet{Kunth:2003}.}
\label{fig:haro11_overlay}
\end{center}
\end{figure*}

Haro~11 (ESO~350-IG~038) is a well-known member of this galaxy class, having been studied in detail in X-ray, UV, optical and infrared wavelengths \citep[e.g.][]{Bergvall:2000,Ostlin:2001,Kunth:2003,Hayes:2007,Adamo:2010,Cormier:2012}.  It is a massive BCG that consists of three main knots (Fig.~\ref{fig:haro11_overlay}), each containing numerous super star clusters with ages between 1 and 40~Myr, and high present star-formation rates (SFR) of $\sim$22~\Msol/yr \citep{Adamo:2010}.   The stellar mass was estimated as $10^{10}$~\Msol\, by \citet{Ostlin:2001}, whereas the total gas mass was derived by \citet{Bergvall:2000} to be $2\times10^9$~\Msol.  An intriguing aspect of Haro~11 is that it is a \Lya\, emitting galaxy and can be considered as both a local (albeit faint) analogue of the high redshift Lyman Break Galaxies (LBGs) or a rather bright Lyman Alpha Emitter \citep{Hayes:2007,Leitet:2011}.  Moreover, it is also in the rare class of objects that are Lyman continuum emitters and its IR luminosity of $\sim2\times10^{11}$~\Lsol\, classes it as a luminous IR galaxy (LIRG) \citep{Adamo:2010}. 

This paper focuses on another interesting aspect of Haro~11 -- its classification as a `high N/O' galaxy.  The distribution of nitrogen as a function of metallicity in star-forming galaxies (SFGs) has often been a cause for debate, especially at intermediate 
metallicities (7.6$\leq$ 12+log(O/H) $\leq$8.3).  Within this metallicity range the 
N/O abundance ratio indicates that nitrogen behaves neither as a primary element 
(i.e. N/O independent of O/H) nor as a secondary element (i.e. N/O 
proportional to O/H); instead a large scatter is observed \citep[see e.g.][their fig.~11]{Lopez-Sanchez:2010a}.  With a nitrogen excess factor of $\sim$6 for its metallicity having been reported for Haro~11 \citep{Izotov:1999}, this galaxy belongs to a small subset of high N/O galaxies highlighted by \citet{Pustilnik:2004}, several of which have been the focus of our integral-field spectroscopic (IFS) observations \citep{James:2009,James:2010,James:2013a}, aimed at understanding the source of such abundance peculiarities.  In the past, long-slit chemical abundance studies have suggested that a connection exists between anomalously high N/O ratios and WR stars, based on their simultaneous detection.  However, spatially resolved abundance studies are showing that this is not a one-to-one relationship \citep[e.g.][]{Monreal-Ibero:2012}.  IFS observations of high N/O galaxies have distinct advantages. By mapping the electron temperatures (\elt) one avoids measuring integrated and therefore likely inappropriately flux-weighted line ratios, enabling more accurate chemical abundances to be computed \citep[e.g.][]{Kobulnicky:1999,James:2010,James:2013a,Pilyugin:2012}. Secondly, the data allows one to isolate regions of enrichment, and thirdly, these regions can be related to environmental factors such as WR-populations, large scale outflows, mergers etc. \citep[]{Walsh:1989,Lopez-Sanchez:2007,James:2009,Perez-Montero:2011,Monreal-Ibero:2012}.  

Haro~11 is known to be a complex system.  Its dynamics base on the \ha\ emission have been studied by \citet{Ostlin:2001}, who reported a multi-component velocity field and a perturbed morphology.  These have been interpreted as signatures of a merger between a low-mass, evolved system and a gas-rich component.  \citet{Bergvall:2002} detected strong emission from WR stars in the spectral region around \heii\, $\lambda$4686 that suggests that Haro~11 may harbour a population of such stars, although to date this population has not been studied in detail. 

In this paper we present integral field spectroscopy observations which afford us a new spatiokinematic `3-D' view of Haro~11. The aims  are three-fold; firstly to map the multiple gas components, secondly to perform a chemodynamical study, i.e. determine the range of physical properties and chemical compositions for each emission line velocity component \citep[e.g.][]{EV:1992,James:2009,Hagele:2012,Amorin:2012}, and thirdly, to spatially relate any chemical anomalies to environmental factors. The adopted distance for Haro~11, along with several other of its properties are listed in Table~\ref{tab:properties}.

\begin{table}
\begin{center}
\begin{footnotesize}
\caption{General Properties of Haro~11}
\begin{threeparttable}
\begin{tabular}{l|cc}
\hline
Property & Value & Reference \\
\hline
RA (J2000) & 00h36$'$52.5$''$ & NED\tnote{a}\\
Dec (J2000) & -33$\deg$33$'$19$''$ & NED\tnote{a}\\
$z$ & 0.020558 & this work \\
Distance / Mpc & 83.6 & this work\tnote{b}\\
Velocity / \kms\, & 6146$\pm$17 & this work\\
Stellar mass / \Msol & $\sim$10$^{10}$ & \citet{Adamo:2010} \\
Gas mass / \Msol & $2\times10^9$ & \citet{Bergvall:2000}\\
Cluster masses / \Msol & $10^3$--$10^6$ & \citet{Adamo:2010}\\
Present SFR / \Msol\,yr$^{-1}$ & 22$\pm$3 & \citet{Adamo:2010} \\
$L_{\sc IR}$ / \Lsol & $1.9\times10^{11}$ & \citet{Bergvall:2000}\\
\hline
\label{tab:properties}
\end{tabular}
\begin{tablenotes}
\item[a] NASA/IPAC Extragalactic Database
\item[b] Derived for a Hubble constant of H$_o=$73.5\,km~s$^{-1}$~Mpc$^{-1}$ \citep{Bernardis:2008}.
\end{tablenotes}
\end{threeparttable}
\end{footnotesize}
\end{center}
\end{table}

\section{Observations and Data Reduction}
\label{sec:obs_red}

\begin{table*}
\begin{center}
\begin{footnotesize}
\caption{FLAMES Argus observing log}
\begin{tabular}{ccccccc}
\hline
Date &Grism & Wavelength Range   &Exp. time & Avg. Airmass & FWHM seeing & Standard Star\\
 & & (\AA) & (s) & & (arcsec) & \\
\hline\hline
03/07/2009 & LR1 & 3620--4081 & 3${\times}$218 & 1.03  & 0.45 & Feige56 \\
.. & LR2 & 3964--4567 & 4$\times$235 & 1.02 & 0.55 & LTT7987\\
.. & LR3 & 4501--5078 & 3$\times$218 & 1.02 & 0.48 & Feige56\\
19/07/2009 & LR6 & 6438--7184 & 4$\times$600 & 1.02 & 0.64 & Feige110\\
20/07/2009 & LR1 & 3620--4081 & 4${\times}$218 & 1.16  & 0.71 & HR5501 \\
.. & LR2 & 3964--4567 & 8$\times$235 & 1.13 & 0.59 & Feige110\\
.. & LR3 & 4501--5078 & 6$\times$218 & 1.06 & 0.67 & Feige110\\
23/07/2009 & LR1 & 3620--4081 & 9${\times}$218 & 1.08  & 0.59 & Feige110 \\
.. & LR2 & 3964--4567 & 8$\times$235 & 1.05  & 0.60 & Feige110\\
.. & LR3 & 4501--5078 & 6$\times$218 & 1.04 & 0.68 & Feige110\\
24/07/2009 & LR6 & 6438--7184 & 4$\times$600 & 1.08 & 0.79 & LTT1020\\
\hline\label{tab:obs}
\end{tabular}
\end{footnotesize}
\end{center}
\end{table*}
Observations of Haro~11 were obtained with the \textit{Fibre Large Array Multi Element Spectrograph}, FLAMES \citep{Pasquini:2002} at Kueyen, Telescope Unit 2 of the 8.2\,m VLT at ESO's Paranal observatory, in service mode on the dates specified in Table~\ref{tab:obs}.  Observations were made with the Argus Integral Field Unit (IFU), with a field-of-view (FoV) of 11\farcs5$\times$7\farcs3 and a sampling of 0\farcs52/lens.  In addition to science fibres, Argus has 15 sky-dedicated fibres that simultaneously observe the sky and that were arranged to surround the IFU FoV.  The positioning of the IFU aperture for Haro~11 is shown overlaid on \textit{HST}-ACS/WFC F435W imaging in Fig.~\ref{fig:haro11_overlay}.

Four different low-resolution (LR) grisms were utilised:  LR1
($\lambda$$\lambda$3620-4081, 24.7$\pm$0.2~\kms\, FWHM resolution), LR2 
($\lambda$$\lambda$3964-4567, 24.9$\pm$0.1~\kms), LR3
($\lambda$$\lambda$4501-5078, 24.7$\pm$0.2~\kms) and LR6 
($\lambda$$\lambda$6438-7184, 22.2$\pm$0.3~\kms). This enabled us to cover all the important emission lines needed for an optical abundance analysis, as detailed in Section~\ref{sec:mapping}.  The seeing (0.45--0.79\arcsec), airmass and exposure time for each data set are detailed in Table~\ref{tab:obs}.  In addition to the science frames, continuum and ThAr arc lamp exposures as well as spectrophotometric standard star observations were obtained (also detailed in Table~\ref{tab:obs}).

The data were reduced using the girBLDRS pipeline (Blecha \& Simond 2004)\footnote{The girBLDRS pipeline is provided by the Geneva Observatory 
http://girbldrs.sourceforge.net} in
a process which included bias removal, localisation of fibres on the
flat-field exposures, extraction of individual fibres, wavelength
calibration and rebinning of the ThAr lamp exposures, and the full
processing of science frames which resulted in flat-fielded,
wavelength-rebinned spectra. The pipeline recipes `biasMast', `locMast',
`wcalMast', and `extract' were used \citep[cf.][]{Tsamis:2008}.
The frames were then averaged using the `imcombine' task of IRAF\footnote{IRAF is
distributed by the National
Optical Astronomy Observatory, which is operated by the Association of 
Universities for Research in Astronomy} which also
performed the cosmic ray rejection. The flux calibration was performed within IRAF using
the tasks \textsc{calibrate} and
\textsc{standard}.  Spectra of the spectrophotometric stars quoted in
Table~\ref{tab:obs} were individually
extracted with girBLDRS and the spaxels containing the stellar emission were summed up
to form a 1D spectrum. The sensitivity function was determined using 
\textsc{sensfunc} and this was subsequently applied to the combined
Haro~11 science exposures. The sky subtraction was performed by averaging the 
spectra recorded by the sky fibres and subtracting this spectrum from that 
of each spaxel in the IFU. Custom-made scripts were then used to convert 
the row by row stacked, processed CCD spectra to data cubes.  This resulted 
in one science cube per grating, i.e. four science cubes for the source
galaxy.

Observations of an object's spectrum through the Earth's atmosphere
are subject to refraction as a function of wavelength, known
as differential atmospheric refraction (DAR). The direction of DAR
is along the parallactic angle at which the observation is made.
Following the method described in James et al. (2009), each reduced data
cube was corrected for this effect using the algorithm created by \citet{Walsh:1990}.

\section{Mapping of Line Fluxes and Kinematics}
\begin{table}
\begin{center}
\begin{scriptsize}
\begin{tabular}{l|r@{$\pm$}lr@{$\pm$}lr@{$\pm$}lr@{$\pm$}|}
\hline
Integrated Spectrum	&	\multicolumn{2}{c}{$F$($\lambda$)}			&\multicolumn{2}{c}{$I$($\lambda$)}			\\
\hline
\foii\, $\lambda$3727	&	126.48	&	2.66	&	136.59	&	7.19	\\
\foii\, $\lambda$3729	&	151.45	&	3.16	&	163.56	&	8.59	\\
\fneiii\, $\lambda$3868	&	34.45	&	1.16	&	36.93	&	2.14	\\
H8+\hei\, $\lambda$3888	&	20.81	&	1.40	&	22.28	&	1.83	\\
\hei\, $\lambda$4025	&	1.59	&	0.11	&	1.70	&	0.14	\\
\fsii\, $\lambda$4068	&	2.05	&	0.11	&	2.19	&	0.15	\\
\hd\, 	&	23.21	&	0.48	&	24.51	&	1.22	\\
\hg\,	&	43.63	&	0.92	&	45.32	&	2.18	\\
\foiii\, $\lambda$4363	&	2.31	&	0.18	&	2.40	&	0.21	\\
\hei\, $\lambda$4471	&	3.55	&	0.11	&	3.65	&	0.19	\\
\ffeii\, $\lambda$4658	&	1.76	&	0.15	&	1.79	&	0.17	\\
\hb\,	&	100.00	&	2.87	&	100.00	&	4.80	\\
\foiii\, $\lambda$4959	&	115.17	&	3.42	&	114.34	&	5.46	\\
\foi\, $\lambda$6364	&	2.12	&	0.06	&	1.95	&	0.07	\\
\fnii\, $\lambda$6548	&	11.92	&	0.29	&	10.83	&	0.38	\\
\ha\,	&	308.67	&	6.51	&	281.06	&	9.50	\\
\fnii\ $\lambda$6584	&	45.42	&	0.99	&	41.22	&	1.40	\\
\hei\, $\lambda$6678	&	2.86	&	0.07	&	2.58	&	0.09	\\
\fsii\, $\lambda$6716	&	25.28	&	0.60	&	22.81	&	0.79	\\

c(\hb) &\multicolumn{4}{c|}{		0.13	$\pm$	0.02				}	\\
F(\hb) $\times10^{14}$ erg s$^{-1}$ cm$^{-2}$ &\multicolumn{4}{c|}{		37.8078	$\pm$	0.77				}	\\
\hline
\end{tabular}
\caption{Haro~11 fluxes and de-reddenened line intensities (relative \hb\ $=$ 100) for the integrated spectrum of Haro~11.  Line fluxes were extinction-corrected using the c(\hb) value shown at the bottom of the table, calculated from the relative \ha, \hb\, and \hg\, fluxes.  The line intensities listed here were used for summed-spectra \elt\, and \eld\, diagnostics and ionic abundance calculations listed in Table~\ref{tab:haro11_avg_map}.}
\label{tab:haro11_sum}
\end{scriptsize}
\end{center}
\end{table}

%

\label{sec:mapping}
\begin{figure*}
\includegraphics[scale=.70]{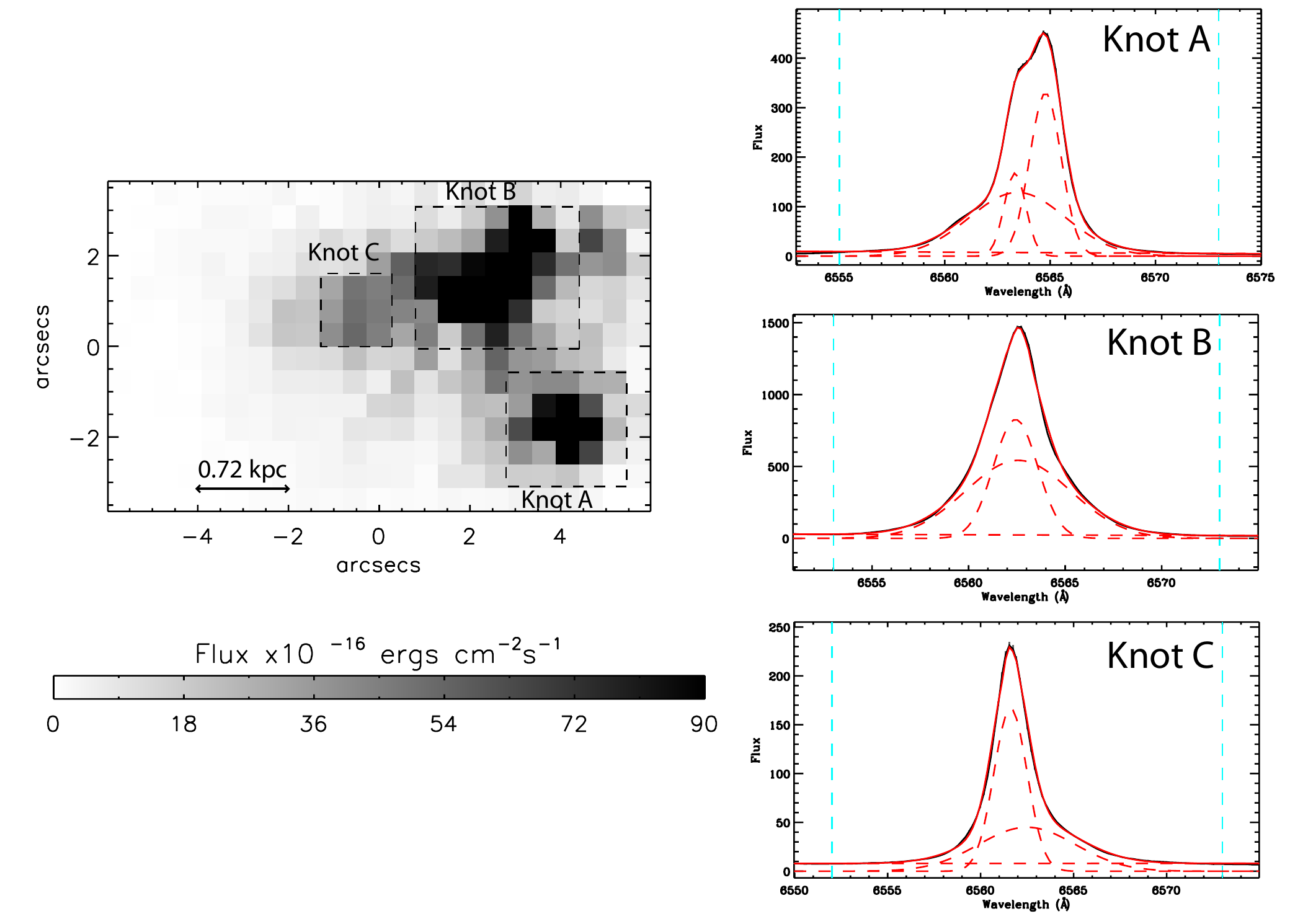}
\caption{\hb\, emission per 0\farcs52$\times$0\farcs52~arcsec$^2$ spaxel. \textit{Left - } Three regions were chosen across Haro~11 whose integrated spectra were extracted for analysis (labelled as Knots A, B and C, after the nomenclature of \citet{Kunth:2003}).  \textit{Right - } The  \ha\, line profile integrated over each region is shown along with a three-component Gaussian fit.  North is up and east is to the left.}
\label{fig:haro11_SFR}
\end{figure*}

The distribution of H~{\sc i} Balmer line emission across Haro~11, which is indicative of
current massive star formation activity, was used to delineate 3 emission line regions surrounding the main knots of emission apparent in the HST ACS image (Fig.~\ref{fig:haro11_overlay}). Our analysis of Haro~11 is based on properties averaged over each of these three areas whose boundaries are displayed on the \ha\, map in Figure~\ref{fig:haro11_SFR} and are labeled as Knots A, B, and C following the nomenclature of \citet{Kunth:2003}.  

Knots B and C on Fig.~\ref{fig:haro11_overlay} correspond to unresolved super-star-clusters, with Knot B containing the youngest cluster population of mass $>$ 8 $\times$ 10$^6$ M$_{\odot}$ (Adamo et al. 2010). Knot C is the brightest UV source in Haro 11 probably due to ongoing massive star formation activity (Hayes et al. 2007). Knot A on the other hand is resolved to a sparse population of clusters. Both Knots B and C are ringed by dust lanes and Knot B appears almost fully surrounded by a dusty spiral arm inclined to the plane of the sky.
Not only the surface brightness, but also the composite structure of the lines varies considerably across Knots A, B, and C (Fig.~\ref{fig:haro11_SFR}).

\begin{figure*}
\includegraphics[scale=1.0]{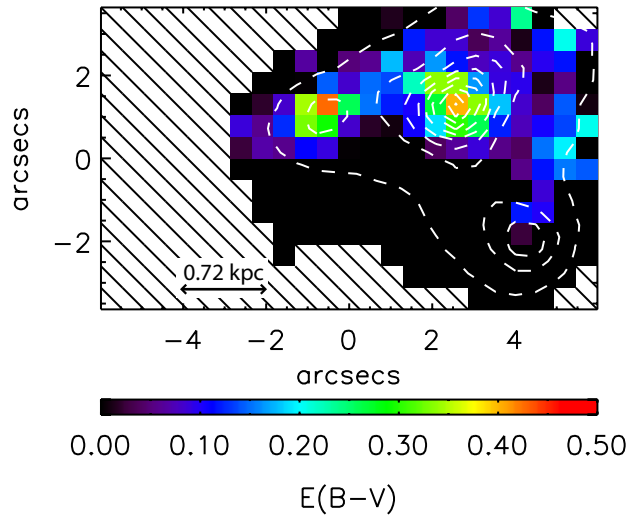}
\caption{Map of E(B-V) for the integrated flux of the emission lines. \ha\, contours are overlaid in white.  North is up and east is to the left.}
\label{fig:haro11_red}
\end{figure*}

The high spectral resolution and signal-to-noise (S/N) ratio of the data allow the identification of multiple velocity components in the majority of emission lines seen in the 300 spectra of Haro~11 across the IFU aperture.  Following the automated line-fitting procedures outlined in \citet[][and references therein]{James:2013a} and likelihood ratio methods described by \citet[][their Appendix A]{Westmoquette:2011}, we were able to find the optimum number of Gaussians required to fit each observed profile.
S/N ratio
maps (used for making S/N cuts) were first made for
each emission line, taking the ratio of the integrated intensity
of the line to that of the error-array produced by the FLAMES pipeline\footnote{The uncertainties output by the pipeline include uncertainties from the flat-fielding and the geometric 
correction of the spectra and not just photon-noise.}, on 
a spaxel-by-spaxel basis. Single or multiple Gaussian profiles were then
fitted to the emission lines, restricting the minimum FWHM to be the instrumental width
(see Section~\ref{sec:obs_red}). Suitable wavelength limits were defined for
each line and continuum level fit. 

\begin{figure*}
\includegraphics[scale=.70]{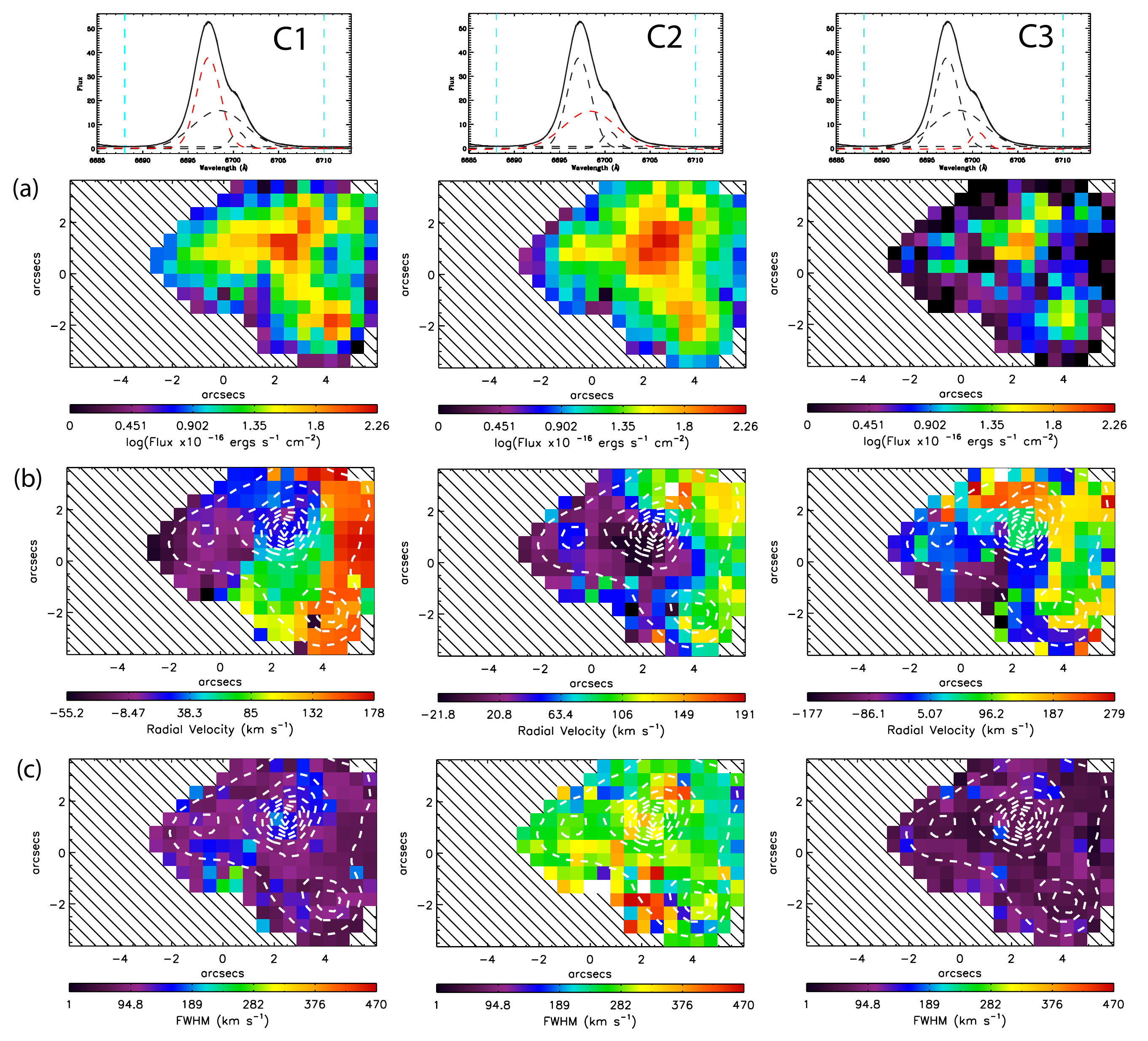}
\caption{Maps of Haro~11 in the \ha\, velocity components C1-C3.  The first row shows an example three-component \ha\, line profile.  The highlighted component indicates to which component the maps in each column correspond; a) logarithmic flux (in units of $10^{-16}~ergs~s^{-1}~cm^{-2}~arcsec^{-2}$); b) radial velocity (relative to the heliocentric systemic velocity of 6146~\kms); c) FWHM corrected for the instrumental PSF.  See text for details.  North is up and east is to the left.}
\label{fig:haro11_ha}
\end{figure*}

Most of the high S/N emission line profiles (i.e. all Balmer lines, \foii~$\lambda\lambda$3727, 3729, \foiii~$\lambda$4959\footnote{\foiii~$\lambda$5007 is not within the wavelength range of this data on account of its redshift.},\fnii~$\lambda$6584, \fsii~$\lambda\lambda$~6716, 6731) were optimally fitted with a narrow Gaussian  (FWHM$<$200~\kms, component C1 hereafter), an underlying broad
Gaussian component (FWHM 200-300~\kms, C2), and a third narrow Gaussian (C3 hereafter) which can
appear either red- or blue-shifted with respect to C1.   With respect to the properties of each Gaussian component (i.e. FWHM and/or velocity), we consider them to arise from gas of different physical conditions.  Towards the outer galactic regions the lines are optimally fitted with only the C1 and C2 components.  A few lines (e.g. \hei~$\lambda$4026) are optimally fitted with a single, narrow component (C1).  Figure~\ref{fig:haro11_ha} shows the complex profile of the \ha\, line, along with maps of its constituent velocity components.  

Tables \ref{tab:Haro11-1}--\ref{tab:Haro11-3} list measured FWHMs and observed and de-reddened fluxes for the fitted Gaussian components of the detected emission lines in the three regions defined in Fig.~\ref{fig:haro11_SFR}.  The fluxes are for summed spectra over each SF knot and are quoted relative to the flux of the corresponding \hb\, component.  In addition, table~\ref{tab:haro11_sum} lists the observed and de-reddened integrated line fluxes (i.e. C1+C2+C3) from summed spectra over the entire galaxy.  All fluxes were corrected for reddening using the galactic reddening law of \citet{Howarth:1983} using c(\hb) values derived from the \ha/\hb\, and \hg/\hb\, line ratios of their corresponding components, weighted in a 3:1 ratio, respectively, in conjunction with theoretical Case B ratios from \citet{Storey:1995}.  A Milky Way reddening of $E(B-V)=$ 0.011~mag, in the direction to Haro~11 is indicated by the extinction maps of \citet{Schlegel:1998}, corresponding to c(\hb)=0.016.  Total c(\hb) values applicable to the galaxy and its individual emission line components in each main SF knot are listed in their respective tables.

 An extinction map was also derived using the \ha/\hb\, and \hg/\hb\, emission line ratio maps and is shown in Fig.~\ref{fig:haro11_red}.  This was computed using the integrated line fluxes (i.e. C1$+$C2$+$C3). The distribution in E(B-V) is in agreement with that found in the \Lya\, study by \citet{Hayes:2007}, where the extinction appears to peak at the locations of Knot B and C whereas Knot A shows little or no extinction. As pointed out by \citet{Hayes:2007}, it is interesting that the source of \Lya\, emission within Haro~11 is located in Knot C, whereas one would expect a low level of extinction to be needed to allow the escape of \Lya\, photons.  This indicates that dust is not the major regulatory factor governing the \Lya\, morphology in Haro~11, which was found by \citet{Hayes:2007} to be driven more by the \hi\, and kinematical structure.  This is similar to what is found in many other \Lya\, emitting galaxies \citep[e.g.][and references therein]{Kunth:1998,Mas-Hesse:2003,Hayes:2010}.


\subsection{\ha\, maps}
In the light of \ha\, (Fig.~\ref{fig:haro11_ha}a) Haro~11 consists of three main regions: a central body located at Knot B ($\alpha = 00^{\rm h}\, 36^{\rm m}\, 52.41\pm0.1^{\rm s}$, $\delta =-33^\circ\,33'\,16.82\pm0.1''$, J2000), and two separated peaks, Knot C ($\sim$3\farcs1, $\sim$1.2~kpc to the east) and Knot A ($\sim$3\farcs2, $\sim$1.3~kpc to the south).  Knots A and B are clearly evident in all 3 line components, with Knot B appearing most pronounced in C2.  Knot C appears as a lower surface brightness feature in C1 and C2, projected against a fainter background connecting it to Knot B.  The inter-knot background appears fainter in C3.  In integrated light (C1+C2+C3), the \ha\, flux of $\sim1.06\times10^{-12}$~ergs~cm$^{-2}$~s$^{-1}$, in reasonable agreement with \citet{Ostlin:1999} and \citet{Vader:1993}. 

The \ha\, radial velocity maps of Haro~11 are shown in Fig.~\ref{fig:haro11_ha}b (relative to the systemic velocity).  The C1 velocity component shows a velocity gradient orientated in the NE-SW direction, ranging from $\sim-$50 to +100~\kms, and an increase up to $\sim$200~\kms\, along the western edge. A tentative rotation axis can be defined based on component C1 which is at an angle of 90 deg to the dusty spiral lane surrounding Knot B in Fig.\,~1. In contrast to this, the velocity gradient of C2 ranges from $\sim-$20 to +100~\kms and extends in the NW-SE direction across Knot~B, i.e. orthogonal to that of C1. This peculiar velocity structure is also highlighted by \citet{Ostlin:2001}, who suggest that this central region contains `a counter-rotating disc or high velocity blobs', which we are tracing with component C2 and is centred on Knot B. This counter-rotating disc may also be evident as dusty filaments to the east of Knot B cutting across the spiral that is ringing it (Fig\,~1). Overall, and in accordance with the conclusions of \citet{Ostlin:2001}, the combined velocity distributions of C1 and C2 suggest that the center of Haro~11 is not dynamically relaxed, while the outer velocity field shows a rather slow rotation. Component C3 shows a more patchy velocity distribution but there is evidence for a wide-angled bipolar outflow oriented approximately along the north-south direction centred on Knot B. The overall velocity distribution of C3 could also be interpreted as the lower S/N ratio signature of geometrically thin shells created by stellar outflows and supernova remnants projected against the background emission.  

FUSE observations of Haro~11 by \citet{Grimes:2007} report similar findings to the \ha\, velocity maps presented here, albeit in one-dimension and relative to a slightly higher systemic velocity of 6180~\kms. UV absorption lines show strong absorption features (with FWHMs of $\sim$300~\kms) blue-shifted by 100~\kms\, relative to the host galaxy systemic velocity (rather equivalent to our C2 component) and an additional, weaker high-velocity outflow $+$200-280~\kms\, away from the galaxy (our C1 component). Interestingly, when comparing the equivalent widths, the higher ionization UV lines are stronger in the high-velocity outflow than the low-velocity absorption features, which Grimes et al. (2007) suggest could be due to shocks between the outflow and the ambient material. 

The maps of \ha\, FWHM (Fig.~\ref{fig:haro11_ha}c) show minimal structure in each of the three main velocity components.  FWHMs range from $\sim$80--180~\kms\, for C1, $\sim$180--450~\kms\, for C2 and $<$90~\kms\, for C3.  \citet{Ostlin:2001} observe a maximum FWHM of $\sim$270~\kms\, for the emission lines within Haro~11, which agrees well with the average FWHM of 283$\pm$5~\kms seen for C2 throughout the main body of the galaxy. The largest FWHM value for C2 is observed directly east of Knot A coinciding with a super star cluster positioned at the end of a bright arc of emission extending southwards from Knot B (Fig.~\,1).  This region spatially aligns with a region of blueshifted C3 emission and could thus be a further signature of outflow activity originating in Knot B.

\subsection{Forbidden Lines}
\begin{figure*}
\includegraphics[scale=0.6]{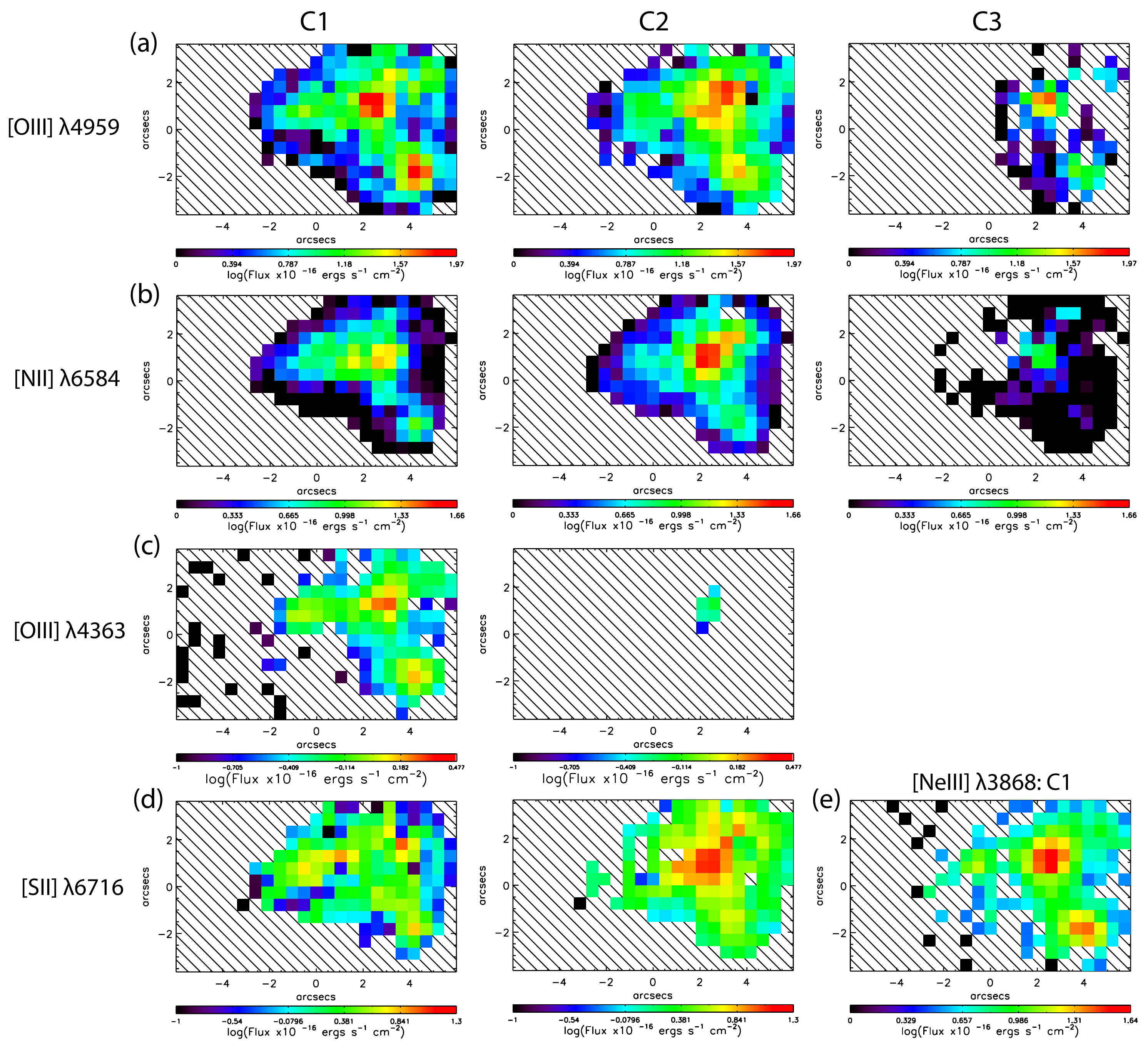}
\caption{Dereddened intensity maps of Haro~11 in \foiii~$\lambda$4959, 
\fnii~$\lambda$6584, \foiii~$\lambda$4363 and  \fneiii~$\lambda$3868 for the separate
velocity components C1, C2 and C3. \foiii~$\lambda$4363 is shown after being re-binned by
1.5$\times$1.5 spaxels and re-mapped onto the original grid (see the text for details).
Only spaxels with S/N$\geq$10 are shown for \foiii~$\lambda$4959 and \fnii~$\lambda$6584,
$\geq$5 for  \fneiii~$\lambda$3868 and $\geq$1 for \foiii~$\lambda$4363.  North is up and
east is to the left.}
\label{fig:haro11_fmaps}
\end{figure*}
Emission line maps of Haro~11 in the light of the \foiii, \fnii, \fsii, and \fneiii\, lines are shown in Fig.~\ref{fig:haro11_fmaps}.  The morphology of \foiii~$\lambda$4959 is similar to that of \ha, with the peak in C2 emission being more extended than that of C1.   Despite the low S/N of the \foiii~$\lambda$4363 line, we were able to map the C1 velocity component throughout Haro~11, with peaks that correlate with those in \ha\, along with the C2 component within Knot B.  
\begin{figure}
\includegraphics[scale=0.8]{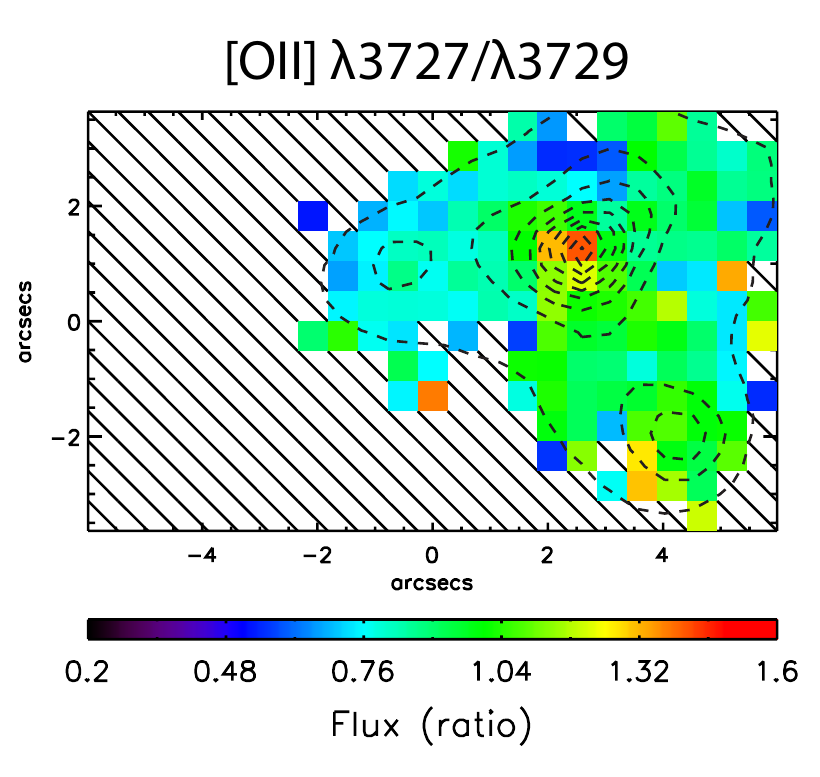}
\caption{De-reddened emission line maps of Haro~11 in \foii~$\lambda$3726/$\lambda$3729, with integrated light \ha\, emission line contours overlaid in dashed line. North is up and east is to the left.}
\label{fig:haro11_O2}
\end{figure}

We show Haro~11 in the light of \fsii~$\lambda$6716 emission in Fig.~\ref{fig:haro11_fmaps}d, in the C1 and C2 velocity components.  Unlike the C1 components of the \fnii\ and \foiii\, emission lines, \fsii~$\lambda$6716 does not show clear peaks in surface brightness, but instead remains relatively constant throughout the main body of Haro~11.  In comparison, the C2 velocity component of \fsii~$\lambda$6716 peaks strongly in Knot B with a morphology akin to \ha~C2 emission (Fig.~\ref{fig:haro11_ha}a - central panel).  Figure~\ref{fig:haro11_fmaps}e shows Haro~11 in the light of \fneiii\,~$\lambda$3868, which was only present in a narrow (C1) component, that has three clear peaks at the three main knots of star-formation.

Figure~\ref{fig:haro11_O2} shows Haro~11 in the light of the electron density (\eld) sensitive doublet ratio, \foii~$\lambda\lambda$3726/3729.  A small amount of structure is visible; the ratio peaking at the location of Knot B, which suggests a region of increased \eld.  The broad component of the \foii~$\lambda\lambda$3726, 3729 doublet cannot be reliably decomposed as their relatively small rest wavelength separation is blurred out by its large width.  
The \fsii~$\lambda\lambda$6716, 6731 doublet on the other hand is unusable as at the redshift of Haro~11 the latter line falls in a spectral region of strong telluric oxygen absorption.

\section{Diagnostics: Electron Temperature and Density}
The dereddened \foiii\ ($\lambda$4959)/$\lambda$4363 and \foii\, $\lambda$3726/$\lambda$3729
intensity ratios were used to determine electron temperatures and electron densities throughout Haro~11, respectively.  
The \elt\ and \eld\ values were computed by inputting the \foiii, and \foii\, intensity ratios
into \textsc{iraf's} {\tt ZONES} task in the {\tt NEBULA} package. This task derives both the temperature and the density by making simultaneous use of temperature- and density-sensitive line ratios. Atomic transition probabilities for \opp\, and \op\, were taken from \citet{Wiese:1996} whilst collision strengths were taken from \citet{Lennon:1994} and \citet{McLaughlin:1993}, respectively.

\begin{figure*}
\includegraphics[scale=1.0]{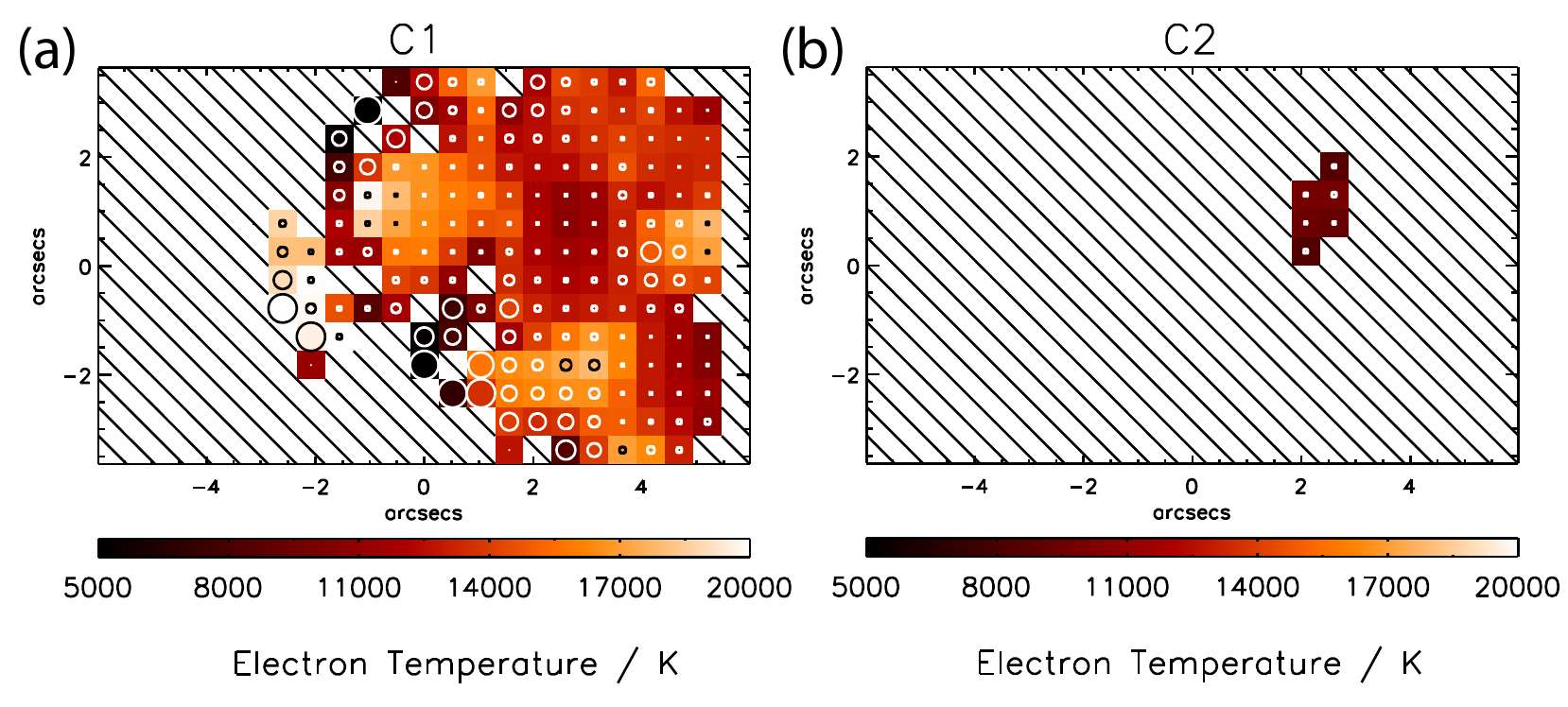}
\caption{Electron temperature map of Haro~11 for velocity components C1 and C2. Overlaid circles represent the size of the uncertainty in \elt\, within each spaxel.  North is up and east is to the left.}
\label{fig:haro11_te}
\end{figure*}

\begin{figure}
\includegraphics[scale=1.0]{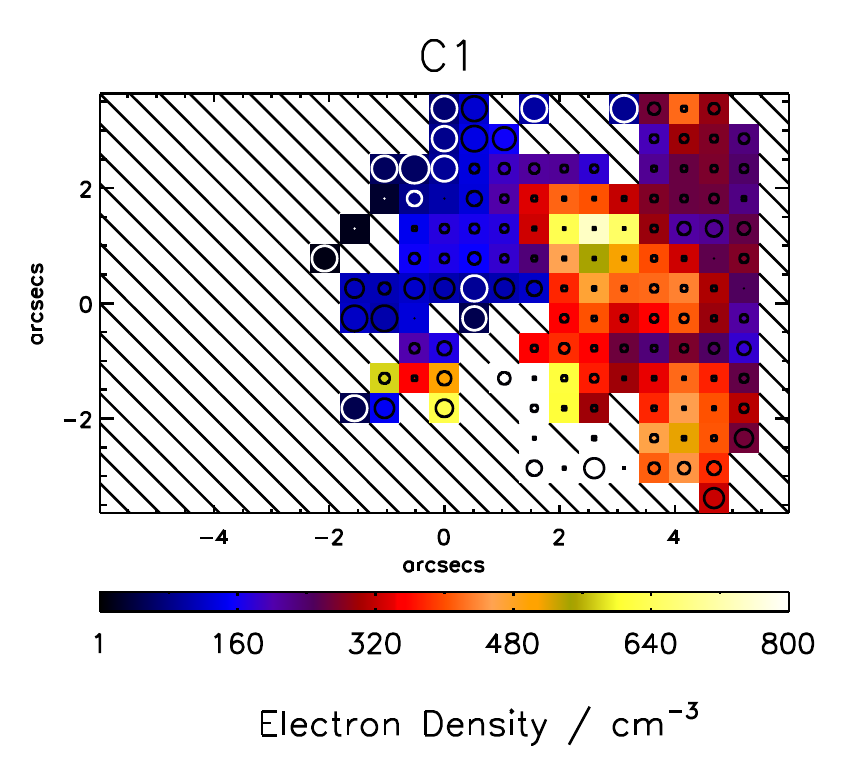}
\caption{Electron density map of Haro~11 for velocity components C1 and C2. Overlaid circles represent the size of the uncertainty in \elt\, within each spaxel.  North is up and east is to the left.}
\label{fig:haro11_ne}
\end{figure}

\elt\, maps computed for line components C1 and C2 are shown in Fig.~\ref{fig:haro11_te} and the \eld\, map for C1 is shown in Fig.~\ref{fig:haro11_ne}. The per spaxel S/N ratio of the \elt\ maps is in the range 10--20.  The error-weighted mean values for each of the Knots (i.e. map-based averages across apertures defined in Fig.~\ref{fig:haro11_SFR}) are listed in Table~\ref{tab:haro11_avg_map}, along with uncertainties corresponding to the standard-deviation across each aperture.  Fig.~\ref{fig:haro11_te}a shows evidence for a small amount of \elt\, variation throughout Haro~11.  For component C1 Knots A and B show temperatures of $\sim$12--13,000~K whilst Knot C shows $\sim$16,000~K.  The broad line component in Knot B has a lower temperature of $\sim$9,000~K.  

Electron density variations are also apparent and denser areas are correlated with Knots A and B which show densities of $\sim$400~cm$^{-3}$ whereas Knot C is less dense with $\sim$150~cm$^{-3}$.  The low \eld\, seen in Knot C may help explain the escape of \Lya\, photons within this region \citep{Hayes:2007}. In Knot C, \citet{Hayes:2007} found that the \Lya/\ha\, ratio agrees with the theoretical case B recombination value, \Lya/\ha\,=8.7 \citep{Osterbrock}, indicating that the \Lya\, photons from this region are scattered just a few times before escaping. Such direct escape would be possible from a low-density, dust-poor nebula bordering on the optically thin Case A, however, as we discussed in the introduction to Section 3, the dust content is probably not the determinant factor in this case; the lower density may be the cause.  As suggested by the models of \citet{Neufeld:1991}, \Lya\, photons will suffer less attenuation than radiation that is not resonantly scattered when they exist in a multi-phase medium, i.e. dusty gas clouds embedded within a low-density inter-cloud medium.  Our \eld\, values based on the C1 \foii~$\lambda$3726/$\lambda$3729 ratio is not in agreement with that of \citet{Guseva:2012}; this may be because their \fsii\, diagnostic is affected by telluric absorption of the $\lambda$6731 component.


\section{Chemical Abundances}
\begin{figure}
\includegraphics[scale=0.9]{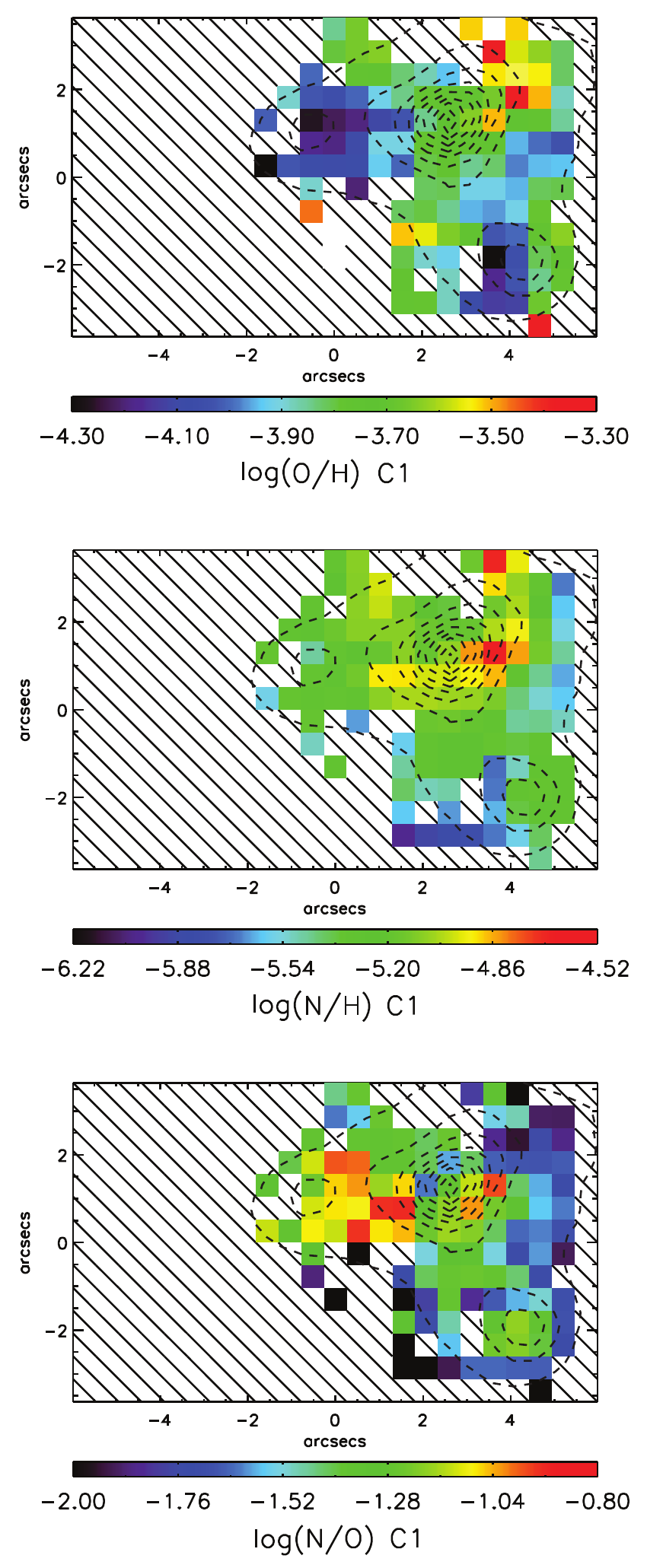}
\caption{ Elemental nitrogen, oxygen and N/O abundance maps for velocity component C1, with integrated light \ha\, emission line contours overlaid in dashed line. North is up and east is to the left.}
\label{fig:haro11_abund}
\end{figure}

Ionic abundance maps relative to H$^+$ were created for the \np, \op, \opp, \nep\, and \Sp\, ions, using the $\lambda\lambda$6584, 3727+3729, 4959, 3868 and 6717 lines respectively.  Examples of flux maps used in the derivations of these ionic abundance maps can be seen in Figure~\ref{fig:haro11_fmaps}.  
Abundances were calculated using the \textsc{zones} task in \textsc{IRAF}, using the respective \elt\, and \eld\, maps described above, with each FLAMES spaxel treated as a distinct `nebular zone' with its own set of physical conditions.  The IRAF abundance results are corroborated by comparing with abundance maps computed via the programme \textsc{equib} (originally written by I. D. Howarth and S. Adams), which uses different atomic data tables; we found that typically the differences in the abundance ratios are within 5\%.

Ionic nitrogen, neon and sulphur abundances were converted into N/H, Ne/H and S/H abundances using ionisation correction factors (ICFs) from \citet{Barlow:1994}.  The O/H abundance was
obtained by adding the \opp /\hp\, and \op /\hp\, abundance maps (i.e. assuming that
the O$^{3+}$/\hp\ ratio is negligible). Since
the \fsiii~$\lambda$6312 line was not within the wavelength range of our 
data, the \Spp/\hp\, abundance was estimated using the empirical relationship between the
S$^{2+}$ and S$^+$ ionic fractions from the corrected equation
A38 of \citet{Barlow:1994}, namely S$^{2+}$/S$^+$ = $4.677 \times (O^{2+}/O^+)^{0.433}$.
The S/H results should therefore be considered with caution, but they are not at variance with those obtained using the ICF prescriptions of Izotov et al. (2006). 
O/H, N/H and log(N/O) abundance maps for line component C1 are shown in 
Fig.~\ref{fig:haro11_abund} and are
described below.  Abundances for the three regions were derived from taking 
error-weighted averages across the regions in the abundance maps shown in Fig.~\ref{fig:haro11_abund}.  These map-based spatial averages are listed in Table~\ref{tab:haro11_avg_map} and correspond to the `average values' described the following section.  We were unable to derive ICFs for the C2 line component without maps of \foii-C2, however, averages over the ionic abundance maps are listed in Table~\ref{tab:haro11_avg_map}.

\subsection{Maps}\label{sec:abund}
The elemental oxygen abundance map shown in the top panel of Fig.~\ref{fig:haro11_abund} displays a single peak whose location correlates with Knot B (the brightest of the knots in terms of \ha\, flux).  Both Knots A and C reveal a decreased oxygen abundance at the location of their peak in \ha\, flux.  Slight abundance variations are seen across Haro~11, with a decreasing oxygen abundance as one moves away from Knot B, with a average metallicity of 12+log(O/H)$=$ 8.25$\pm$0.03, decreasing to 8.09$\pm$0.03 and 7.80$\pm$0.04 for Knots A and C, respectively. There is, therefore, a distinct inhomogeneity in oxygen abundance across Haro~11, with Knot B having a higher oxygen abundance by upto $\sim$0.4~dex.  Whilst the result for Knot B is in relatively good agreement with \citet{Guseva:2012} who derived 12+log(O/H)$=$ 8.33, our abundance for Knot C is somewhat lower than their 8.10 due to our higher temperature (they derive $\sim$11,500~K for Knot C).  We are, however, in agreement with the value 12+log(O/H) $=$ 7.9 obtained by \citet{Bergvall:2002} in a $4''\times4''$ aperture over the brightest region of Haro~11 in the visual region.  Whilst we cannot comment on the C2 elemental abundance of oxygen, Table~\ref{tab:haro11_avg_map} shows that the \opp\ C2 component abundance in Knot B is approximately twice that of its narrow C1 component.


The middle panel of Fig.~\ref{fig:haro11_abund} shows the N/H abundance across Haro~11.  The peak in abundance is located within Knot B, with an average N/H ratio of $9.38\pm0.02\times10^{-6}$, which decreases to $\sim4.88\times10^{-6}$ in Knots A and C.  In contrast to oxygen, there is no distinct decrease in nitrogen abundance in Knot C and as a result this knot shows a peak in N/O, which can be seen in the bottom panel of Fig.~\ref{fig:haro11_abund}.  Whilst Knots A and B display relatively `normal' log(N/O) levels for their respective metallicities ($\sim$--1.4 and --1.3, \citet[see][their figure 11]{Lopez-Sanchez:2010a}), Knot C has an average log(N/O) of --1.12, $\sim$0.5~dex higher than the expected value for 12+log(O/H)=7.80.  In fact, the map of log(N/O) shows that this peak actually reaches $\sim$-$1.0$ to $-$0.8 at the boundary between Knots B and C (as can be seen by the \ha\, contours).  Therefore, the factor of $\sim$3 excess reported by \citet{Pustilnik:2004} is seen by us within Knot C only.  We investigate the cause behind this anomalously high ratio in Section~\ref{sec:Nenrich}.  

Again, we cannot draw firm conclusions on the N/H or N/O abundance within the broad velocity component gas. However, as a rough estimate, if we apply the average ratio of O$^+$/O$^{2+}$=1.57 seen in the C1 component of Knot B to the O$^{2+}$/H$^+$ value in C2, we would obtain O$^{+}$/H$^+ = 2.43\times10^{-4}$ and a log($N^+/O^+$) ratio of $-$1.07. This would provide some evidence for the existence of a broad-line region within Knot B that has a higher O/H abundance than the narrow-line component, but a similar N/O ratio.

\citet{Guseva:2012} derive log(N/O) values of $-$0.92 and $-$0.79 for Knots B and C, respectively; whilst the value for Knot C is in agreement with the maps derived here, the value for Knot B is somewhat higher.  We interpret this as being a result of their using  ratios from integrated line profiles.

The Ne/H distribution in Haro~11 (not shown) follows that of nitrogen in that a peak in abundance is seen in Knot B with Ne/H $=$ 1.0$\times10^{-4}$, whilst Knots A and C have an abundance ratio of $\sim0.4\times10^{-4}$.  The average Ne/O ratios are all $\sim$0.4~dex higher than the expected range of log(Ne/O) $\sim-0.9$ to $-0.4$ \citep{Lopez-Sanchez:2010a}.  Derivations of Ne/H were also made using the neon ICF of \citet{Izotov:2006}, with only the Ne/O ratio for Knot B being $\sim$0.1~dex higher than the expected value. The S/H distribution across Haro~11 (also not shown) is relatively constant for the C1 gas component, with each region displaying 
averages within the range of $\sim$(3--6) $\times$ 10$^{-7}$.  The S/O ratio for Knot C is in agreement with the average reported range 
of log(S/O)$\sim -1.9$ to $-1.5$ \citep{Izotov:2006}  for blue compact galaxies, whilst Knots A and B are $\sim$0.2--0.4~dex lower than the expected values.

\subsection{Comparison of spatially resolved and global spectra abundances}
Also listed in Table~\ref{tab:haro11_avg_map} are the \elt, \eld, ionic and elemental abundances derived from the galaxy integrated spectra (using the line fluxes listed in Table~\ref{tab:haro11_sum}).  Following the methodology of \citet{James:2013a}, this set of results  allows us to assess if integrated galaxy spectra can reliably represent the physical properties of the ionized ISM.  In a comparison between spatially-averaged and integrated-spectra results for UM~448, \citet{James:2013a} found that whilst \emph{relative} abundance ratios of heavy elements can be reasonably well obtained from integrated spectra and/or long-slit observations, the ionic and elemental abundances relative to hydrogen are not well reproduced.  \citet{James:2013a} suggested that this may be because the former properties do not have a very strong sensitivity to \elt\, biases and/or, in contrast to integral field spectroscopy, integrated properties derived from flux-averaged methods cannot resolve abundance and temperature variations and sites of enrichment are being averaged out.

The present results suggest that whilst the N/O and Ne/O values are in agreement with the spatially-averaged values, the ionic and elemental abundance ratios relative to hydrogen are not.  The \elt\, from the integrated spectrum is in excellent agreement with the $\lambda$4959-flux-weighted mean \elt\, across the three regions (10,850~K), suggesting that the integrated \foiii\, ratio may have been more heavily weighted by areas of stronger \foiii~$\lambda$4959 emission.  Indeed, the average \elt\, for Knot B (which has the strongest \foiii\, emission) from the C1 and C2 maps, would be $\sim$10,550~K  - in agreement with the integrated-spectrum \elt.  These discrepancies suggest that values derived from long-slit observations may be prone to luminosity-weighted fluxes and resultant inaccurate ratios, compounded by the fact that large apertures can include a mixture of gas with different ionization conditions and metal content. Such effects can lead to significant biases in analyses based on integrated spectra of galaxies both in the local and the high-redshift universe.  These biases may be of comparable magnitude to the effects of
temperature or density variations \citep[e.g.][]{Peimbert:1969,Rubin:1989,Viegas:1994} often invoked in nebular abundance analyses.

\subsection{Helium Abundances}
The \hep/\hp\, abundances for Knots A and B are listed in Table~\ref{tab:haro11_Heflux}. The \hei\ lines in Knot C are abnormally weak relative to \hb\ probably due to the effects of underlying stellar absorption and hence a reliable He abundance cannot be derived for this area. Summed spectra over each knot and velocity component (i.e. C1+C2+C3) were used rather than map-based averages because of the low per spaxel S/N ratio of the lines. Fluxes were de-reddened using the c(\hb) value for its corresponding knot. Helium abundances were calculated using the error-weighted line fluxes listed in Table~5 and adopting the map-based average \elt\, and \eld\, for each knot (listed in Table~\ref{tab:haro11_avg_map}). The Case B \hei\, emissivities of \citet{Smits:1996} were adopted, correcting for the effects of collisional excitation using the formulae in \citet{Benjamin:1999}.  The presence of neutral helium within the \hp\ zone has been accounted for using the ICF of \citet{Peimbert:1992}. No substantial amounts of \hepp\ are expected to be present. 




\begin{table*}
\begin{center}
\begin{footnotesize}
\begin{tabular}{|l|c|c|c|c|c}
\hline   
 & \multicolumn{3}{c|}{C1} & \multicolumn{1}{c|}{C2} & Summed Spectra\\
\cline{2-6}
& Knot A & Knot B & Knot C & Knot B  & \\
\hline   

\elt & 12600$\pm$270 & 12100$\pm$230 & 16600$\pm$480 & 9000$\pm$400                                                                                               &	10,340	$\pm$	440	\\       		
\eld & 410$\pm$20 & 370$\pm$10 & 150$\pm$15 & ---                                                                                                                  &	180	$\pm$	100	\\                   	
\noalign{\vskip4pt}                                                                                                                                               				                        

O$^+/H^+$ $\times10^5$ &         5.64$\pm$        0.85 &        10.92$\pm$        1.28 &         3.91$\pm$        0.39     &   ---&                                      1.34	$\pm$	0.07    \\ 
O$^{++}/H^+$ $\times10^5$ &       6.78$\pm$        0.84 &         6.92$\pm$        0.69 &         2.45$\pm$        0.22     &   15.38$\pm$        2.71               &   1.19	$\pm$	0.06     \\ 
O/H $\times10^4$ &              1.24$\pm$        0.24 &         1.78$\pm$        0.27 &         0.64$\pm$        0.08      & ---&                                       2.53	$\pm$	0.18    \\
12+log(O/H) &                   8.09$\pm$        0.20 &         8.25$\pm$        0.15 &         7.80$\pm$        0.13     &  ---&                                       8.40	$\pm$	0.07    \\ 
Z/\Zsol &                       0.24$\pm$        0.05 &         0.35$\pm$        0.05 &         0.12$\pm$        0.02     &  ---&                                       0.49	$\pm$	0.04   \\ 
\noalign{\vskip4pt}                                                                                                                                                		    
N$^+/H^+$ $\times10^6$ &         2.19$\pm$        0.22 &         5.74$\pm$        0.43 &         3.00$\pm$        0.21     &   20.28$\pm$        2.74 &                   8.16	$\pm$	0.28   \\ 
ICF(N) & 2.20 & 1.63 & 1.63 & ---                                                                                         &                                                    1.89	        \\
N/H $\times10^6$ &              4.82$\pm$        1.28 &         9.38$\pm$        1.95 &         4.88$\pm$        0.88      &   ---&                                     15.45	$\pm$	1.49     \\ 
log(N$^+$/O$^+$) &              -1.41$\pm$        0.08 &        -1.28$\pm$        0.06 &        -1.12$\pm$        0.05     &   ---&                                      -1.21	$\pm$	0.03    \\ 
\noalign{\vskip4pt}                                                                                                                                                		                                                                               
Ne$^{++}/H^+$ $\times10^5$ &     2.62$\pm$        0.34 &         3.93$\pm$        0.48 &         1.36$\pm$        0.13      &                                              3.75	$\pm$	0.22    \\ 
ICF(Ne) & 1.83 & 2.58 & 2.59 & ---                                                                                         &                                                   2.12	        \\
Ne/H $\times10^5$ &             4.79$\pm$        1.27 &        10.14$\pm$        2.23 &         3.54$\pm$        0.66      & ---&                                         7.94	$\pm$	0.83    \\ 
log(Ne$^{2+}$/O$^{2+}$) &        -0.41$\pm$        0.14 &        -0.25$\pm$        0.12 &        -0.25$\pm$        0.10     &         ---&                                 -0.50	$\pm$	0.03     \\ 
\noalign{\vskip4pt}                                                                                                                                                    		              
S$^+/H^+$ $\times10^7$ &         3.37$\pm$        0.32 &         4.64$\pm$        0.34 &         6.36$\pm$        0.49      &  18.47$\pm$        2.48 &                   8.79	$\pm$	0.55    \\
ICF(S) & 1.06 & 1.02 & 1.02 & ---                                                                                           &                                                   1.04	         \\
S/H $\times10^7$ &              21.71$\pm$        5.35 &        22.93$\pm$        4.44 &        31.25$\pm$        5.19      &    ---&                                     4.97	$\pm$	0.26   \\ 
log(S$^+$/O$^+$) &              -1.76$\pm$        0.14 &        -1.89$\pm$        0.11 &        -1.31$\pm$        0.09     &     ---&                                    -2.18	$\pm$	0.04    \\

\hline   

\end{tabular}
\caption{Ionic and elemental abundances for Haro~11, with the corresponding \elt\, and \eld\, used in their derivation.  Columns 1--4 refer to abundances derived from maps, weight-averaged over regions defined in Fig.~\ref{fig:haro11_SFR}.   All ionisation correction factors (ICFs) are taken from \citet{Barlow:1994}.}
\label{tab:haro11_avg_map}
\end{footnotesize}
\end{center}
\end{table*}

\section{WR stars and N-enrichment}\label{sec:Nenrich}
As discussed previously, a connection is thought to exist between galaxies with high N/O ratios and the presence of Wolf-Rayet (WR) stars \citep[e.g.,][and references therein]{Brinchmann:2008}.  Two examples are NGC~5253 \citep[][and references therein]{Lopez-Sanchez:2007} and Mrk~996 \citep{James:2009}, where spatially-resolved studies showed regions of enhanced N/O that correlate with a strong WR population.  However, the relationship may not be clear-cut.  Cases also exist where galaxies display N-overabundance over areas too large for WR-stars to pollute, suggesting that other mechanisms may be responsible \citep{Perez-Montero:2011,James:2013a}.

\begin{figure}
\includegraphics[scale=.45]{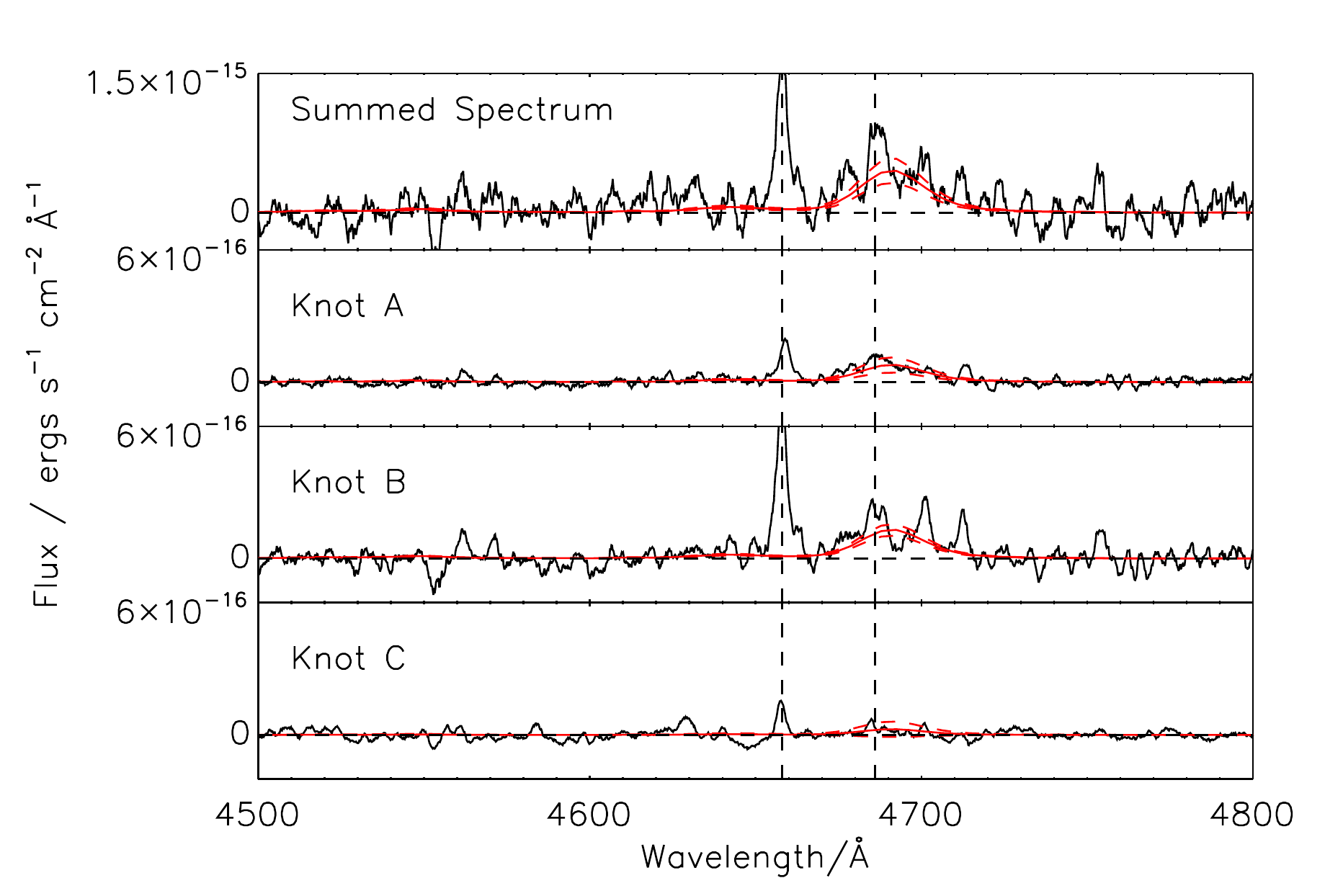}
\caption{Sections of continuum-subtracted FLAMES IFU spectra, showing the blue `WR 
bump' region within summed spectra over the entire 
galaxy and separate aperture regions 1, 2 and 3 (see 
Fig.~\ref{fig:haro11_SFR}).  We detect WR emission 
features in the spectra summed over the entire galaxy and also each of the individual aperture region spectra.  Overlaid are the estimates of 5200$\pm$1500, 900$\pm$400, 1500$\pm$300 and 300$\pm$400 WN stars, for the entire galaxy and Knots A, B and C, respectively, using the templates of \citet{Crowther:2006}.  Vertical dashed lines indicate the location of [Fe~{\sc iii}]
4658~\AA\, and a weak, narrow, \heii\, emission line at 4686~\AA.  The dashed horizontal lines in the plots show the zero 
flux levels.}
\label{fig:WRspec}
\end{figure}

\begin{figure}
\includegraphics[scale=.9]{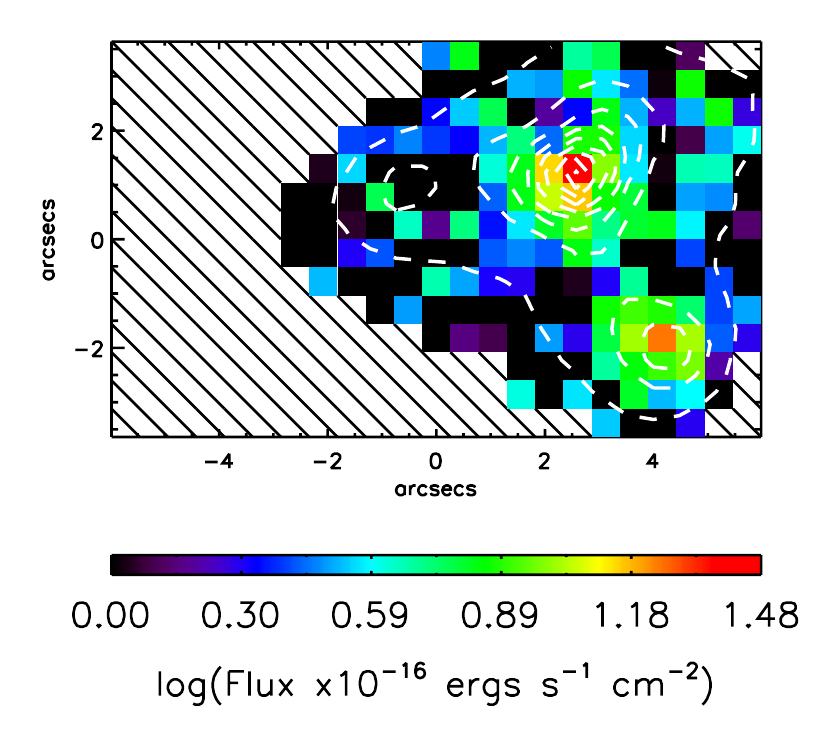}
\caption{Emission map of the blue WR feature, showing the distribution of WR stars throughout Haro~11.  Overlaid dashed contours represent the \ha\, integrated flux. North is up and east is to the left.}
\label{fig:WRmap}
\end{figure}

The presence of WR features within the spectra of Haro~11 was reported by \citet{Bergvall:2002}. However, to-date, no attempt has been made to map or quantify the size of this population.  The FLAMES-IFU spectra also show a broad WR stellar feature around 4690~\AA, attributable to a mixture of late-type WN (WNL) stars, as seen in Fig.~\ref{fig:WRspec}. A map of the WR feature is shown in Fig.~\ref{fig:WRmap} (made by integrating the flux across the feature), where the WR emission is seen to extend throughout Knots B and A, with only a minimal detection in Knot C.  

When searching for WR features one must consider the spatial width of the extraction aperture, which can sometimes be too large and thereby dilute weak WR features with continuum flux \citep{Kehrig:2008,Lopez-Sanchez:2008,James:2010,Perez-Montero:2011}.  We therefore also examined the spectra summed over each individual knot (Fig.~\ref{fig:haro11_SFR}), which are shown in Fig.~\ref{fig:WRspec}.   In order to determine the number of WR stars from the strength of the 4640-4690~\AA\ feature for the summed spectra over Knots A, B and C, we followed the procedure described in James et al. (2012). This makes use of an LMC WR (WN5--6) spectral template from \citet{Crowther:2006} to fit the observed WR broad feature using Monte Carlo to estimate the errors.

It is thus estimated that 900 $\pm$ 400, 1500 $\pm$ 300 and 300 $\pm$ 400 WN stars are present in Knots A, B and C respectively. A total population of 5200 $\pm$ 1500 WN stars is estimated applying the same procedure for the integrated spectrum of the entire galaxy. The difference between this and the total number of WR stars from the regional spectra suggests that there may be WR stars located outside knots of star-formation, similar to the situation in the nearby BCG NGC~5253 \citep[e.g.][and references therein]{Monreal-Ibero:2010}. 



\begin{figure}
\includegraphics[scale=.50]{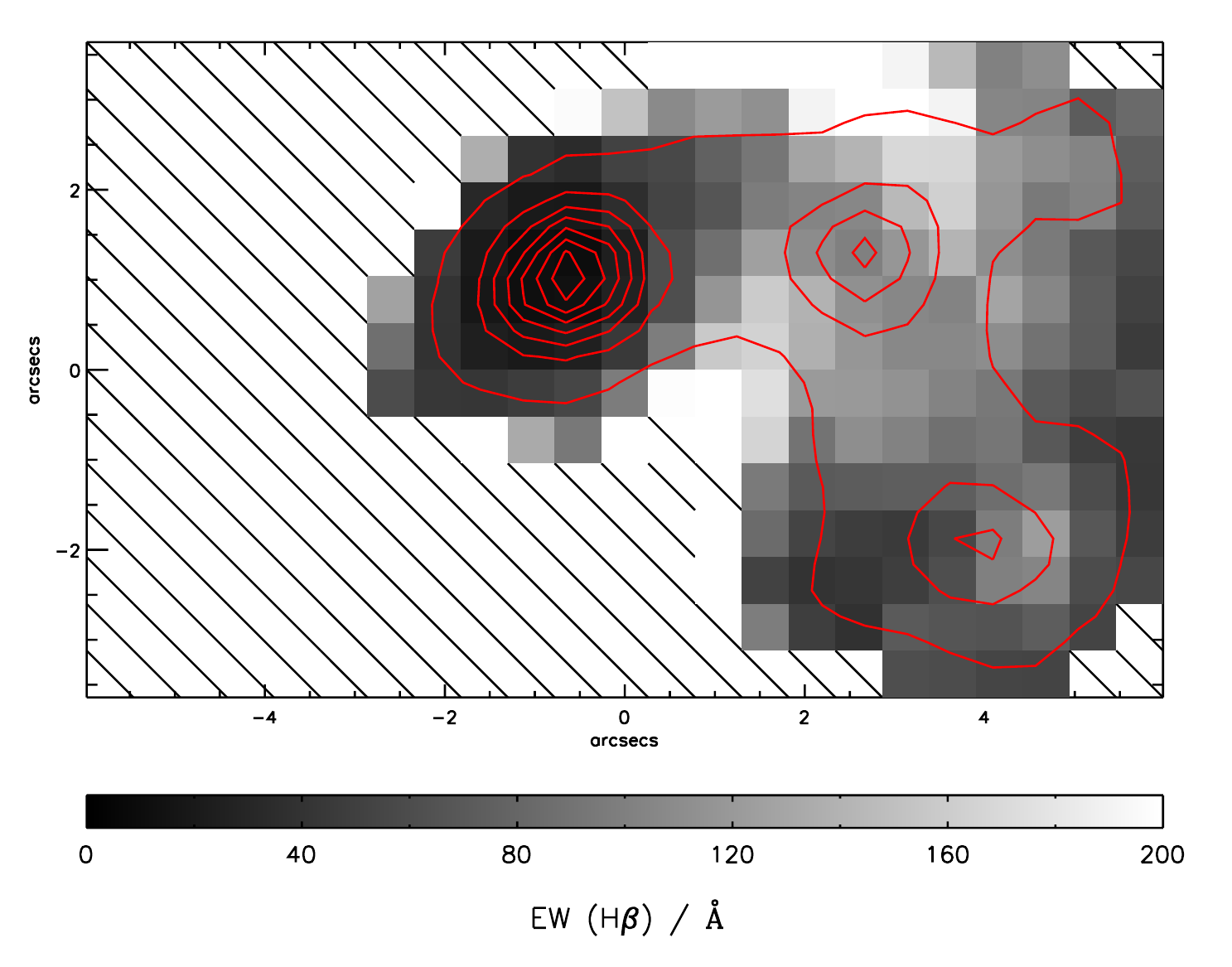}
\caption{Map of the equivalent width of \hb\, across Haro~11.  Overlaid red lines are 
contours of \hb-region continuum.   North is up and east is to the left.}
\label{fig:haro11_EW}
\end{figure}

Due to the short WR phase of a star's evolution, another way to assess if they are present is to estimate the current age of the ionizing stellar population.  This can be achieved by assessing the equivalent-width (EW) of \hb\, at a given metallicity.  A map of $EW$(\hb) is shown in Fig.~\ref{fig:haro11_EW}. Two regions of increased $EW$ are present, located at Knots B and A.  Knot C, on the other hand, shows a distinct decrease in $EW$.  Following the method outlined in \citet{James:2010}, we can use this map, in conjunction with the metallicity map described in Section~\ref{sec:abund}, to estimate the age of the most recent star-forming episodes throughout Haro~11, by comparing the regional observed average EW(\hb) values with those predicted by the spectral synthesis code STARBURST99 \citep{Leitherer:1999}.  We utilize the same model parameters described in \citet{James:2010} and obtain average stellar ages of 4.9$\pm$0.4, 4.3$\pm$0.5 and 7.4$\pm$0.3~Myr for Knots A, B and C, respectively.  These ages are in relatively good agreement with \citet{Adamo:2010} who derive ages of 3.9 and 9.5~Myr for Knots B and C, respectively from modeled spectral energy distributions of the star clusters.  According to \citet{Leitherer:1999}, typically an instantaneous starburst shows WR features at ages of $\sim$3--6 Myr for metallicities of Z = 0.004--0.008, similar to those seen in Haro~11.  Thus the stellar population ages of Knots A and B are consistent with hosting WR stars, whilst Knot C appears too old to do so.  

Due to their high mass, the number ratio of WR stars to O stars within a galaxy can be used to provide constraints on the high-mass end of the initial mass function (IMF).  Following the methodology outlined in \citet{James:2009}, an estimation of the number of O-type stars within each knot was made using L(\hb), $N_{\rm WR}$, and the number of O7V-equivalent stars \citep[e.g.][]{Kunth:1981,Osterbrock:1982}, after applying the appropriate correction factor for the proportion of other O star subtypes, $\eta_0(t)$, for the metallicity of each knot.  The emission rate of ionising photons, $Q_0^{obs}$, are 1.01, 3.20 and 0.34 $\times10^{53}$ photons~s$^{-1}$ for Knots A, B, and C, respectively, yielding average O-star populations of size $\sim$45,900, $\sim$101,700 and $\sim$25,800.  As a result, we find $N_{\rm WR}/N_{\rm WR+O}$ ratios of 0.020, 0.014 and 0.012 for Knots A--C.  These results are in agreement with the predictions of \citet{Schaerer:1998} for Knots A and C, but for an IMF with $\alpha=3.0$ \citep[see][their figure 13]{Lopez-Sanchez:2010b}, i.e. slightly steeper than the Salpeter-like IMF with $\alpha=$2.35.  However, the ratio for Knot B calls for an even steeper IMF.  One must bear in mind that here we are assuming $N_{\rm WR}=N_{\rm WNL}$ and, as \citet{Lopez-Sanchez:2010b} point out, the inclusion of WCE stars may increase the $N_{\rm WR}/N_{\rm WR+O}$ ratio sufficiently to reconcile our results with the standard Salpeter-IMF.

By comparing the map of N/O (Fig.~\ref{fig:haro11_abund}) and WR emission (Fig.~\ref{fig:WRmap}) we are thus faced with two intriguing situations.  Firstly, for Knot C, we have a large area of gas with a relatively high N/O ratio and a small WR population, and secondly, for Knots B and A, we have the reverse situation; regions with relatively low N/O ratio but large WR populations.  In the following we will discuss scenarios that may explain these circumstances (for guidance, we summarise the main properties of each knot in Table~\ref{tab:summary}).
\begin{table}
\begin{center}
\begin{footnotesize}
\caption{Summary of Knot Properties}
\begin{tabular}{cccc}
\hline
Property & Knot~A & Knot~B & Knot~C \\
\hline
\Lya\, emitter & \xmark & \xmark & \cmark \\
WR population & 5200$\pm$1500 & 900$\pm$400 & 300$\pm$400 \\
$N_{WR}/N_{WR+O}$ & $\sim$0.020 & $\sim$0.014 & $\sim$0.012 \\
SF age / Myr & 4.9$\pm$0.4 &  4.3$\pm$0.5 & 7.4$\pm$0.3 \\
Current SFR / \Msol\, yr$^{-1}$ &  0.26 &  0.86 & 0.09  \\
12+log(O/H) & 8.09$\pm$0.20 & 8.25$\pm$0.15 & 7.80$\pm$0.13\\
log(N/O)& -1.41$\pm$0.08 &  -1.28$\pm$0.06 & -1.12$\pm$0.05 \\
\hline\label{tab:summary}
\end{tabular}
\end{footnotesize}
\end{center}
\end{table}

\textit{Knot C:}  The negligible amount of WR emission within this region, combined with the large projected area of high N/O suggests that N-rich winds may not be the cause of the observed $\Delta$log(N/O)$\sim$+0.5~dex excess.  Whilst, this enrichment may be related to the stellar wind ejecta coming from WR stars in neighbouring Knot B, the $\sim$1~kpc separation distance between Knot B's peak in WR emission and the Knot C region of enhanced N/O suggests this is unlikely.  Instead, the excess may be due to other global processes occurring in Haro~11.  For example, the galaxy's unrelaxed dynamics suggests that it is a merger of two bodies.  Indeed, the differing age, \eld, \elt, and metallicity of Knot C compared to the rest of Haro~11 (see Table~\ref{tab:summary}) suggest that it has undergone a separate evolutionary path.  

The decreased oxygen abundance in Knot C indicates that accretion of metal-poor gas might be implicated; some studies suggest that interacting galaxies fall $\sim$0.2~dex below the mass-metallicity relation of normal galaxies due to tidally induced large-scale metal-poor gas inflow to the galaxiesÕ central regions \citep[e.g.][]{Kewley:2006,Michel:2008,Peeples:2009}.  However, the stellar mass of Knot~C \citep[$1.36\times10^{7}$~\Msol][]{Adamo:2010} and its map-derived average metallicity (12+log(O/H)=7.8$\pm$0.13), actually place it in direct agreement with the M--Z relationship at low masses \citep[see][their figure 6]{Zahid:2010}, i.e. alongside low-mass dwarf irregulars from \citep{Lee:2006} and other blue compact dwarf galaxies from \citep{Zhao:2010}.  

Conversely, there may have been an outflow of oxygen-enriched gas due to supernovae \citep[see e.g.][]{VanZee:2006}, rather than an inflow of metal-poor gas.  The effective yield of oxygen and nitrogen may not be in-sync  since the former is formed in high-mass stars, and the latter in intermediate-mass stars, and it is the more recently enriched gas that is ejected via supernova winds \citep[e.g. the case of NGC~1569, ][]{Martin:2002}. Indeed, being 2-3~Myr older than the rest of the galaxy, Knot C may have experienced more SNe explosions than the younger Knots, A and B. This scenario might be more plausible with the aforementioned top-heavy IMF that is required to explain the number ratio of WR/O stars.


Alternatively, the age of Knot C (7.4~Myr) implies that its WR phase has recently been completed, whereas in Knots B and A (both with ages of $\sim$4--5~Myr) the WR phase is still ongoing. This suggests that Knot C has had sufficient time to maximise the injection of N into the local ISM during its WR phase, which may still be relatively undiluted by subsequent mass loss of lower mass stars.  This case is supported by Knot C having a lower current star-formation rate \citep[$\sim$0.09~\Msol\ yr$^{-1}$, as derived from the \ha\, flux listed in Table~\ref{tab:haro11_Heflux} and the][relationship] {Kennicutt:1998}, in comparison to Knot B (0.86~\Msol\ yr$^{-1}$) and Knot A (0.26~\Msol\ yr$^{-1}$).  The situation witnessed here, where the mixed ejecta have cooled but the WR-phase has finished confirms that the relationship between N-enrichment and the existence of WR-features is far from being one-to-one.  In fact, Haro~11 is highly analogous to the nearby `high N/O' BCD NGC~5253, where N-enrichment is not always directly associated with the presence of WR stars \citep[][]{Westmoquette:2012,Monreal-Ibero:2010}.  As more spatially-resolved analyses come to light on  `high N/O objects', it is becoming apparent that the simultaneous detection of both WR-features and enhanced N/O is most probably a function of starburst age and the properties of the surrounding medium.  As proposed by \citet{Westmoquette:2012}, in the two known cases where an increase in N/O \textit{is} coupled with a WR-population, i.e. `Knot 1' in NGC~5253 \citep[][and references therein]{Monreal-Ibero:2012} and the central region of Mrk~996 \citep{James:2009}, we are only able to observe the enriched gas because the high density and pressure of the ISM surrounding these areas may have acted to impede the WR winds, allowing their metal-enriched ejecta to mix with the cooler phases and subsequently become detectable in the warm ($\sim10^4$~K) gas.

\textit{Knots A+B:}
The existence of a strong WR population within these knots, but at normal levels of N/O suggest that we are observing the galaxy at an interesting point in its evolutionary path.  This situation has also been observed in NGC~5253 \citep[see results for `Knot \# 2' of ][]{Monreal-Ibero:2012}, where WR-features were seen in a region of clusters $\sim$1--5~Myr in age, with normal N/O levels for its metallicity.  Here the authors suggest that we are observing a region whose stars are in the process of expelling their processed nitrogen (which begins $\lesssim5$~Myr) and have not had time enough for this material to cool down and mix with the ISM.  Under the assumptions of instantaneous mixing and that the velocity of this gas traces the velocity under which the contamination of extra nitrogen propagates through the ISM, they estimated that the cooling down and mixing of expelled material can last between $\sim$2--8~Myr over an area 40--50~pc in diameter.  For comparison, the areas of Knots A and B are $\sim$0.4 and $\sim$1~kpc$^2$, respectively.  If this injection of extra nitrogen takes place as early as $\sim$2.5~Myr \citep[according to the models of][]{Molla:2012}, the N-enriched material has had only $\sim$2~Myr to be expelled, mixed and diluted in the interstellar medium across these regions.  

In addition to the timing of these observations, we should also consider that our limited spatial resolution may be a factor. Even for nearby 30~Doradus that contains Of- and WR-type stars, nitrogen self-enrichment in that nebula has not been detected with any certainty \citep[]{Rosa:1987}, and the \hii\ region overall shows a normal N/O ratio for its 0.5\,Z$_{\bigodot}$ metallicity \citep[]{Tsamis:2003,Tsamis:2005}.



\begin{table*}
\begin{center}
\begin{footnotesize}
\begin{tabular}{l|r@{$\pm$}lr@{$\pm$}lr@{$\pm$}lr@{$\pm$}lr@{$\pm$}lr@{$\pm$}lr@{$\pm$}lr@{$\pm$}lr@{$\pm$}lr@{$\pm$}}\hline   

	&	\multicolumn{4}{c}{Knot A}					&	\multicolumn{4}{c}{Knot B}										\\
	&	\multicolumn{2}{c}{$F$($\lambda$)}			&	\multicolumn{2}{c}{$I$($\lambda$)	}	&	\multicolumn{2}{c}{$F$($\lambda$)}			&	\multicolumn{2}{c}{$I$($\lambda$)	}	\\
			\hline																						
\hei\, $\lambda$4471	&	3.22	&	0.12	&	3.23	&	0.22	&	3.93	&	0.21	&	4.16	&	0.29		\\
\hei\, $\lambda$6678	&	3.10	&	0.21	&	3.09	&	0.23	&	3.55	&	0.10	&	2.89	&	0.11		\\\noalign{\vskip4pt}     																									
																									
$c$(\hb) &\multicolumn{4}{c|}{		0.01	$\pm$	0.01					}&\multicolumn{4}{c}{	0.27	$\pm$	0.02			}\\
																									
\noalign{\vskip4pt}

He$^+$/H$^+\times10^2$ ($\lambda$4471)&\multicolumn{4}{c|}{		6.53	$\pm$	0.41					}&\multicolumn{4}{c}{	8.31	$\pm$	0.58			}\\    
He$^+$/H$^+\times10^2$ ($\lambda$6678)&\multicolumn{4}{c|}{		8.34	$\pm$	0.62					}&\multicolumn{4}{c}{	7.82	$\pm$	0.28			}\\    

He$^+$/H$^+\times10^2$&\multicolumn{4}{c|}{		7.43	$\pm$	0.54					}&\multicolumn{4}{c}{	7.74	$\pm$	0.30			}\\ 
ICF(He)&\multicolumn{4}{c|}{		1.18							}&\multicolumn{4}{c}{	1.25					}\\

He/H$\times10^2$&\multicolumn{4}{c|}{		8.80	$\pm$	0.64					}&\multicolumn{4}{c}{	9.68	$\pm$	0.38			}\\

\hline   

\end{tabular}
\caption{Regional He/H abundances for Knots A and B from integrated line fluxes (here in units relative to \hb\ $=$ 100) summed over the regions defined in Fig.~\ref{fig:haro11_SFR}.}.  
\label{tab:haro11_Heflux}
\end{footnotesize}
\end{center}
\end{table*}

\section{Summary and conclusions}
We have analysed FLAMES-IFU integral field spectroscopy of Haro~11, a BCG well known for its high star-formation efficiencies and \Lya~emission, making it analogous to the high-redshift LBGs.  We have studied its morphology by creating monochromatic emission line maps.  The galaxy is seen to consist of three star-forming knots (named A, B and C), arranged in the shape of a heart with Knot~B, the region of strongest \ha\, surface brightness in the centre and Knots A and C to the east and south-east, respectively.  Haro~11 exhibits complex emission line profiles, with most lines consisting of a narrow, central component, an underlying broad component, and a third narrow component.  In agreement with previous studies of \ha\, kinematics, this system is not dynamically relaxed and is possibly an ongoing merger. In particular, the contrasting kinematics of the broad and narrow emission line components suggest the presence of a counter-rotating disk within the central region of the galaxy.

For the first time we present 200~pc resolution maps of \elt, \eld, and the abundance ratios of O/H, N/H, Ne/H and S/H across Haro~11. The abundance maps were derived using the direct method of estimating electron temperature from the \foiii~$\lambda$4363/$\lambda$4959 line ratio.  

Variations in oxygen abundance are seen across Haro~11 in the narrow emission component, with a peak in abundance of 12 $+$ log(O/H) $=$ 8.25$\pm$0.15 in Knot B, decreasing to 8.09$\pm$0.23 and 7.80$\pm$0.13 in Knots A and C, respectively.  We measured a difference of $\sim$2000~K between the temperature of the broad and narrow line emission regions in Knot B, but the spectral resolution of our data is not sufficiently high to allow the decomposition of \foii\, in the broad regions. Thus the total oxygen abundance differential between the two gas phases remains unknown.

The low O/H abundance ratio in Knot~C tallies with the increased \elt\ in that region. We detect a high N/O ratio \citep[previously reported for the whole galaxy][]{Izotov:1999} only in Knot C, with $\Delta$log(N/O)$\sim+0.5$~dex, spatially coincident with the decrease in oxygen abundance. This knot also shows a distinctly lower \eld\, in comparison to the rest of the galaxy. This may relate to the previously reported \Lya\, emission observed solely from this region. 

We assess the reliability of abundances derived from galaxy-integrated spectra by comparing those derived from integrated spectra with spatial averages.  Whilst abundances relative to oxygen are consistent within the uncertainties, abundances relative to hydrogen are not. These results illustrate the superior information content of abundances from spatially resolved spectra.

Using the 4686~\AA\, WR signature, we detect evidence of a large WR population in Knots A and B.  No evidence for WR stars was found in Knot~C, although we do measure an enhanced N/O and a decreased O/H ratio.  This leads us to believe that the abundance anomaly in this region may be due to a recently completed WR phase where N remains undiluted by subsequent mass loss of lower mass stars.  The strong WR emission in Knots A+B combined with `normal' N/O levels and a young stellar population ($<$5~Myr) suggest we are observing these knots at an epoch where ejected N-rich material is yet to mix with the warm ionized gas.  Alternatively, even the spatial resolution of these data (0.52$''$ spaxel$^{-1}$), corresponding to $\sim$210~pc at the distance of Haro~11), may be preventing us from detecting small-scale N-enhancements that have already formed. 

To conclude, the IFS data presented in this study show the complexity of Haro~11 as a system; not only is it complicated kinematically but also chemically. In order to better understand the evolution of this system, the inhomogeneous conditions and metallicity of its ISM exposed by our chemodynamical study need to be combined with its star-formation history in models. Observations at even higher resolution are required to constrain the mixing timescales of the ejecta from its large WR population. Successful modeling of an LBG analogue such as this could have a significant impact on our understanding of the chemical homogeneity, ISM mixing, and effect of feedback on galaxy evolution in both the local and high-$z$ universe.

\bibliographystyle{mn2e}
\bibliography{references/references}

\section{acknowledgments}
The authors would like to thank the FLAMES staff at Paranal and Garching for scheduling and taking
these service mode observations [programme 083.B-0336A; PI: B. L. James].  We thank the referee for a careful reading of the manuscript and a constructive report. We appreciate
discussions with Ana Monreal-Ibero regarding theories of N-enrichment from WR-stars.  YGT acknowledges support from a Marie Curie intra-European Fellowship within the 7th European Community Framework Programme
(grant agreement PIEF-GA-2009-236486). BLJ acknowledges support from an STScI-DDRF travel grant used to visit collaborators of this study.  This research made use of the NASA ADS and NED data bases.


\appendix
\section{Regional Observed Fluxes and Line Intensities}
Tables~\ref{tab:Haro11-1}--\ref{tab:Haro11-3} list the observed fluxes and de-reddened line intensities derived from summed spectra over the regions defined in Fig.~\ref{fig:haro11_SFR}.  Fluxes are given for each separate velocity component (C1--C3), along with its de-reddened line intensity and FWHM.
\begin{table*}
\begin{center}
\begin{footnotesize}
\begin{tabular}{l|r@{$\pm$}lr@{$\pm$}lr@{$\pm$}lr@{$\pm$}lr@{$\pm$}lr@{$\pm$}lr@{$\pm$}lr@{$\pm$}lr@{$\pm$}lr@{$\pm$}lr@{$\pm$}lr@{$\pm$}l}
\hline

 Knot A & \multicolumn{6}{c|}{	C1	}										 & \multicolumn{6}{c|}{			C2							}	\\
 & \multicolumn{2}{c|}{		$F$($\lambda$)	}		 & \multicolumn{2}{c|}{	$I$($\lambda$)	}	 & \multicolumn{2}{c|}{	FWHM	}	 & \multicolumn{2}{c|}{	$F$($\lambda$)	}	 & \multicolumn{2}{c|}{	$I$($\lambda$)	}		 & \multicolumn{2}{c|}{	FWHM	}	\\
 \hline
\foii\, $\lambda$3727	&	299.97	&	11.85	&	315.67	&	13.40	&	176.47	&	0.83													\\
\foii\, $\lambda$3729	&	296.16	&	11.73	&	311.66	&	13.26	&	176.36	&	0.83													\\
\fneiii\, $\lambda$3868	&	111.34	&	5.18	&	116.58	&	5.70	&	158.20	&	4.17													\\
H8+\hei\, $\lambda$3889	&	64.33	&	4.24	&	67.32	&	4.55	&	168.31	&	8.94													\\
\hei\, $\lambda$4026	&	3.00	&	0.28	&	3.12	&	0.29	&	126.20	&	10.69													\\
\fsii\, $\lambda$4068	&	2.39	&	0.28	&	2.48	&	0.29	&	124.34	&	14.69													\\
\hd\, 	&	43.23	&	2.65	&	44.82	&	2.83	&	120.78	&	2.68	&	19.37	&	1.63	&	19.58	&	1.69	&	247.11	&	8.02	\\
\hg\,	&	47.49	&	4.10	&	48.70	&	4.26	&	80.30	&	4.40	&	45.58	&	1.84	&	45.91	&	2.05	&	251.43	&	6.14	\\
\foiii\, $\lambda$4363	&	6.15	&	1.63	&	6.30	&	1.68	&	121.07	&	18.29	&	2.28	&	1.49	&	2.27	&	1.48	&	318.69	&	128.11	\\
\hei\, $\lambda$4471	&	6.23	&	1.39	&	6.35	&	1.42	&	117.51	&	11.21 &2.34 &0.99 &2.35 & 1.00 &245.70 &54.33 \\												
\ffeiii\, $\lambda$4657	&	3.50	&	0.46	&	3.54	&	0.47	&	213.39	&	26.67													\\
\hb\,	&	100.00	&	5.53	&	100.00	&	5.60	&	80.28	&	2.20	&	100.00	&	2.21	&	100.00	&	2.78	&	253.78	&	2.57	\\
\foiii\, $\lambda$4959	&	143.94	&	7.60	&	143.25	&	7.66	&	72.61	&	1.81	&	127.54	&	3.36	&	125.70	&	3.01	&	255.08	&	2.23	\\
\foi\, $\lambda$6364	&	2.15	&	0.13	&	2.03	&	0.12	&	164.21	&	7.27													\\
\fnii\, $\lambda$6548	&	6.70	&	0.37	&	6.29	&	0.35	&	62.58	&	2.20	&	10.51	&	0.24	&	10.32	&	0.27	&	276.55	&	5.03	\\
\ha\,	&	317.46	&	13.38	&	298.32	&	12.82	&	81.42	&	0.77	&	288.62	&	4.64	&	283.56	&	5.62	&	263.30	&	0.53	\\
\fnii\ $\lambda$6584	&	24.97	&	1.06	&	23.41	&	1.01	&	74.61	&	1.09	&	28.68	&	0.52	&	28.146&	0.60	&	259.11	&	2.11	\\
\hei\, $\lambda$6678	&	2.85	&	0.31	&	2.66	&	0.29	&	67.35	&	4.69	&	3.00	&	0.27	&	2.95	&	0.27	&	259.27	&	19.96	\\
\fsii\, $\lambda$6716	&	10.84	&	0.61	&	10.13	&	0.59	&	72.78	&	2.19	&	15.37	&	0.35	&	15.08	&	0.38	&	239.06	&	2.35	\\
																									
c(\hb) &\multicolumn{6}{c|}{		0.09	$\pm$	0.01	}		&\multicolumn{6}{c|}{						0.02	$\pm$	0.01		}							\\
F(\hb)$\times$10$^{14}$Â &\multicolumn{6}{c|}{		1.99	$\pm$	0.08	}		&\multicolumn{6}{c|}{						2.73	$\pm$	0.04		}							\\
 \hline
\end{tabular}
\begin{tabular}{l|r@{$\pm$}lr@{$\pm$}lr@{$\pm$}lr@{$\pm$}lr@{$\pm$}lr@{$\pm$}l}
\hline
	&\multicolumn{6}{c|}{C3}											\\
 & \multicolumn{2}{c|}{		$F$($\lambda$)	}	 & \multicolumn{2}{c|}{	$I$($\lambda$)	}		 & \multicolumn{2}{c|}{	FWHM	}	\\
 \hline
\foii\, $\lambda$3727													\\
\foii\, $\lambda$3729													\\
\fneiii\, $\lambda$3868													\\
H8+\hei\, $\lambda$3889													\\
\hei\, $\lambda$4026													\\
\fsii\, $\lambda$4068													\\
\hd\, 													\\
\hg\,	&	43.92	&	8.93	&	44.61	&	9.97	&	58.48	&	5.17	\\
\foiii\, $\lambda$4363													\\
\hei\, $\lambda$4471													\\
\ffeiii\, $\lambda$4657													\\
\hb\,	&	100.00	&	12.14	&	100.00	&	14.54	&	57.76	&	2.51	\\
\foiii\, $\lambda$4959	&	171.58	&	18.76	&	165.52	&	20.61	&	59.28	&	2.15	\\
\foi\, $\lambda$6364													\\
\fnii\, $\lambda$6548	&	5.57	&	0.71	&	5.35	&	0.75	&	49.33	&	3.28	\\
\ha\,	&	288.47	&	27.36	&	276.75	&	30.48	&	62.37	&	1.11	\\
\fnii\ $\lambda$6584	&	12.48	&	1.30	&	11.96	&	1.41	&	42.66	&	1.71	\\
\hei\, $\lambda$6678	&	3.78	&	0.76	&	3.62	&	0.75	&	58.65	&	5.99	\\
\fsii\, $\lambda$6716	&	13.14	&	1.49	&	13.14	&	1.84	&	52.72	&	2.07	\\
													
$c$(\hb) &\multicolumn{6}{c|}{		0.06	$\pm$	0.04		}							\\
$F$(\hb)$\times$10$^{14}$ erg s$^{-1}$ cm$^{-2}$&\multicolumn{6}{c|}{		0.86	$\pm$	0.07		}							\\

\hline

\end{tabular}
\caption{Haro~11 regional fluxes and  de-reddenened line intensities (both relative to \hb\ $=$ 100) and FWHMs (in \kms) for summed spectra over Knot~A (as defined in Fig.~\ref{fig:haro11_SFR}).  Line fluxes were extinction-corrected using the c(\hb) values shown at the bottom of the table, calculated from the relative \ha, \hb\, and \hg\, fluxes.}  
\label{tab:Haro11-1}
\end{footnotesize}
\end{center}
\end{table*}

\begin{table*}
\begin{center}
\begin{scriptsize}
\begin{tabular}{l|r@{$\pm$}lr@{$\pm$}lr@{$\pm$}lr@{$\pm$}lr@{$\pm$}lr@{$\pm$}lr@{$\pm$}lr@{$\pm$}lr@{$\pm$}lr@{$\pm$}lr@{$\pm$}lr@{$\pm$}l}

\hline
 Knot B & \multicolumn{6}{c|}{		C1	}										 & \multicolumn{6}{c|}{				C2						}	\\
 & \multicolumn{2}{c|}{		$F$($\lambda$)	}		 & \multicolumn{2}{c|}{	$I$($\lambda$)	}	 & \multicolumn{2}{c|}{	FWHM	}	 & \multicolumn{2}{c|}{	$F$($\lambda$)	}	 & \multicolumn{2}{c|}{	$I$($\lambda$)	}		 & \multicolumn{2}{c|}{	FWHM	}	\\
\hline
\foii\, $\lambda$3727	&	344.56	&	8.92	&	469.13	&	44.25	&	256.33	&	17.33													\\
\foii\, $\lambda$3729	&	355.21	&	9.19	&	483.63	&	45.61	&	256.11	&	17.32													\\
\fneiii\, $\lambda$3868	&	87.79	&	2.96	&	115.98	&	11.03	&	212.82	&	4.82													\\
H8+\hei\, $\lambda$3889	&	59.32	&	2.76	&	78.09	&	7.82	&	209.83	&	8.22													\\
\hei\, $\lambda$4026	&	4.01	&	0.20	&	5.09	&	0.51	&	172.42	&	7.98													\\
\fsii\, $\lambda$4068	&	4.58	&	0.24	&	5.75	&	0.58	&	188.22	&	8.86													\\
\hd\, 	&	24.43	&	1.31	&	30.42	&	3.07	&	121.96	&	2.80	&	24.06	&	0.60	&	24.76	&	1.72	&	296.04	&	4.70	\\
\feii\, $\lambda$4286 	&	1.69	&	0.32	&	2.00	&	0.41	&	213.13	&	42.98													\\
\hg\,	&	39.50	&	2.51	&	46.04	&	4.75	&	121.30	&	3.00	&	46.56	&	1.23	&	47.50	&	3.19	&	298.60	&	4.44	\\
\foiii\, $\lambda$4363	&	3.60	&	0.66	&	4.17	&	0.83	&	132.66	&	17.21	&	2.69	&	0.57	&	2.68	&	0.57	&	412.75	&	110.16	\\
\hei\, $\lambda$4471	&	4.78	&	0.46	&	5.37	&	0.67	&	122.62	&	6.51	&	3.51	&	0.22	&	3.56	&	0.31	&	384.48	&	28.20	\\
\ffeiii\, $\lambda$4657	&	4.47	&	0.28	&	4.75	&	0.46	&	204.16	&	11.96													\\
\hb\,	&	100.00	&	3.57	&	100.00	&	8.06	&	118.66	&	1.63	&	100.00	&	1.63	&	100.00	&	5.71	&	290.99	&	2.21	\\
\foiii\, $\lambda$4959	&	118.11	&	3.62	&	114.74	&	8.82	&	104.61	&	1.16	&	116.51	&	6.43	&	111.89	&	4.68	&	276.68	&	1.47	\\
\foi\, $\lambda$6364	&	2.06	&	0.20	&	1.47	&	0.16	&	119.18	&	5.77	&	1.99	&	0.10	&	1.91	&	0.12	&	325.55	&	16.60	\\
\fnii\, $\lambda$6548	&	13.61	&	0.61	&	9.28	&	0.62	&	89.60	&	1.55	&	14.41	&	0.28	&	13.71	&	0.58	&	200.66	&	2.06	\\
\ha\,	&	417.83	&	10.80	&	286.85	&	16.06	&	125.02	&	0.41	&	306.78	&	3.67	&	292.02	&	11.54	&	294.78	&	0.41	\\
\fnii\ $\lambda$6584	&	56.97	&	1.59	&	38.59	&	2.17	&	100.03	&	0.70	&	55.54	&	0.69	&	52.77	&	2.06	&	264.94	&	1.21	\\
\hei\, $\lambda$6678	&	2.80	&	0.22	&	1.87	&	0.17	&	91.73	&	3.41	&	3.92	&	0.10	&	3.72	&	0.16	&	242.46	&	6.23	\\
\fsii\, $\lambda$6716	&	15.05	&	0.50	&	9.96	&	0.58	&	97.79	&	1.26	&	25.50	&	0.33	&	24.16	&	0.93	&	253.33	&	1.00	\\
																									
$c$(\hb) &\multicolumn{6}{c|}{		0.52	$\pm$	0.03	}		&\multicolumn{6}{c|}{						0.07	$\pm$	0.02		}							\\
$F$(\hb)$\times$10$^{14}$ erg s$^{-1}$ cm$^{-2}$ &\multicolumn{6}{c|}{		5.83	$\pm$	0.15	}		&\multicolumn{6}{c|}{						11.88	$\pm$	0.14		}							\\

 \hline
\end{tabular}
\caption{Same as for Table~\ref{tab:Haro11-1}, for summed spectra over Knot B (as defined in Fig.~\ref{fig:haro11_SFR}).}
\label{tab:Haro11-2}
\end{scriptsize}
\end{center}
\end{table*}

\begin{table*}
\begin{center}
\begin{scriptsize}
\begin{tabular}{l|r@{$\pm$}lr@{$\pm$}lr@{$\pm$}lr@{$\pm$}lr@{$\pm$}lr@{$\pm$}lr@{$\pm$}lr@{$\pm$}lr@{$\pm$}lr@{$\pm$}lr@{$\pm$}lr@{$\pm$}l}
\hline
Knot C & \multicolumn{6}{c|}{		C1	}										 & \multicolumn{6}{c|}{				C2						}	\\
 & \multicolumn{2}{c|}{		$F$($\lambda$)	}		 & \multicolumn{2}{c|}{	$I$($\lambda$)	}	 & \multicolumn{2}{c|}{	FWHM	}	 & \multicolumn{2}{c|}{	$F$($\lambda$)	}	 & \multicolumn{2}{c|}{	$I$($\lambda$)	}		 & \multicolumn{2}{c|}{	FWHM	}	\\
\hline
\foii\, $\lambda$3727	&	208.00	&	9.85	&	245.20	&	28.29	&	125.59	&	0.72													\\
\foii\, $\lambda$3729	&	263.87	&	12.46	&	311.07	&	35.88	&	125.47	&	0.72													\\
\fneiii\, $\lambda$3868	&	53.77	&	4.66	&	62.38	&	8.40	&	125.33	&	10.34													\\
H8+\hei\, $\lambda$3889	&	19.62	&	4.75	&	22.71	&	5.98	&	81.12	&	21.74													\\
\hei\, $\lambda$4026	&	1.20	&	0.40	&	1.36	&	0.47	&	114.56	&	40.39													\\
\fsii\, $\lambda$4068	&	6.99	&	0.63	&	7.89	&	1.06	&	135.69	&	11.52													\\
\hd\, 	&	37.01	&	1.84	&	41.60	&	4.61	&	99.87	&	1.73													\\
\feii\, $\lambda$4244	&	6.65	&	1.05	&	7.32	&	1.36	&	174.51	&	26.47													\\
\feii\, $\lambda$4286 	&	8.56	&	1.21	&	9.36	&	1.59	&	188.33	&	26.61													\\
\hg\,	&	43.32	&	7.81	&	47.01	&	9.57	&	80.67	&	5.68	&	43.84	&	9.31	&	47.54	&	11.10	&	172.37	&	21.11	\\
\foiii\, $\lambda$4363	&	6.77	&	1.98	&	7.32	&	2.24	&	185.05	&	59.63				\\
\hei\, $\lambda$4471	&	3.86	&	0.39	&	4.10	&	0.56	&	82.76	&	8.15													\\
\ffeiii\, $\lambda$4657	&	5.37	&	0.75	&	5.55	&	0.92	&	185.77	&	26.72													\\
\hb\,	&	100.00	&	6.58	&	100.00	&	10.65	&	82.08	&	2.13	&	100.00	&	9.40	&	100.00&12.74		&	206.87	&	6.83	\\
\foiii\, $\lambda$4959	&	109.67	&	6.29	&	107.99	&	10.79	&	81.02	&	1.60	&	101.58	&	8.18	&	100.04	&	11.63	&	213.87	&	6.95	\\
\foi\, $\lambda$6364	&	3.08	&	0.23	&	2.58	&	0.25	&	111.20	&	6.54													\\
\fnii\, $\lambda$6548	&	11.81	&	1.39	&	9.62	&	1.26	&	67.18	&	3.16	&	17.62	&	1.95	&	14.39	&	1.80	&	147.45	&	7.96	\\
\ha\,	&	355.13	&	16.74	&	290.59	&	21.64	&	97.37	&	0.58	&	357.83	&	23.94	&	293.34	&	26.19	&	286.36	&	1.29	\\
\fnii\ $\lambda$6584	&	50.24	&	2.59	&	40.82	&	3.13	&	78.31	&	1.09	&	50.36	&	3.58	&	41.00	&	3.77	&	202.66	&	4.16	\\
\hei\, $\lambda$6678	&	4.09	&	0.24	&	3.29	&	0.26	&	88.60	&	3.08													\\
\fsii\, $\lambda$6716	&	28.18	&	1.60	&	22.62	&	1.79	&	79.02	&	1.38	&	23.39	&	1.94		&18.81	&	1.89		&191.49		&5.89	\\																									
$c$(\hb) &\multicolumn{6}{c|}{		0.28	$\pm$	0.04	}		&\multicolumn{6}{c|}{						0.28	$\pm$	0.04		}							\\
$F$(\hb)$\times$10$^{14}$ erg s$^{-1}$ cm$^{-2}$ &\multicolumn{6}{c|}{		1.07	$\pm$	0.05	}		&\multicolumn{6}{c|}{						0.81	$\pm$	0.05		}							\\
\hline
\end{tabular}
\caption{Same as for Table~\ref{tab:Haro11-1}, for summed spectra over Knot C (as defined in Fig.~\ref{fig:haro11_SFR}).}
\label{tab:Haro11-3}
\end{scriptsize}
\end{center}
\end{table*}

\bsp

\label{lastpage}

\end{document}